\documentclass[%
 reprint,
 amsmath,amssymb,
 aps,
prb,
]{revtex4-1}
	\usepackage{feynmf}
	\usepackage{graphics}
	\usepackage{epsfig}
	\usepackage{graphicx}
	\usepackage{color}
	\usepackage{comment}
	\usepackage{ulem}

 \newcommand{\nairo}{Na$_2$IrO$_3$}

 \newcommand{\ruclc}{RuCl$_3$}
 
\begin{document}

\title{Challenges in Design of Kitaev Materials: Magnetic Interactions from Competing Energy Scales}% Force line breaks with \\
%\thanks{A footnote to the article title}%

\author{Stephen M. Winter}%
 %\email{winter@physik.uni-frankfurt.de}
\author{Ying Li}
\author{Harald O. Jeschke}
\author{Roser Valent{\'\i}}
\affiliation{Institut fur Theoretische Physik, Goethe-Universitat Frankfurt, 60438 Frankfurt am Main, Germany}

\begin{abstract}
In this study, we reanalyze the magnetic interactions in the Kitaev spin liquid candidate materials Na$_2$IrO$_3$, $\alpha$-RuCl$_3$, and $\alpha$-Li$_2$IrO$_3$ using nonperturbative exact diagonalization methods. These methods are more appropriate given the relatively itinerant nature of the systems suggested in previous works. We treat all interactions up to third neighbours on equal footing. The computed terms reveal significant long range coupling, bond-anisotropy, and/or off-diagonal couplings which we argue naturally explain the observed ordered phases in these systems. Given these observations, the potential for realizing the spin-liquid state in real materials is analyzed, and synthetic challenges are defined and explained. 
\end{abstract}


\maketitle

\section{Introduction}
\subsection{Introduction}

Recently, great interest has emerged in the realization of materials with quantum spin liquid (QSL) ground states, which represent prominent examples of highly entangled and topologically nontrivial phases, and therefore provide a fertile ground for realizing exotic excitations.\cite{balents2010spin} However, stabilizing such states in real systems represents a significant experimental and theoretical challenge. In most cases, classical ordering of spins can only be avoided for significant geometrical frustration, and/or exotic long-range or multi-spin interactions that may not be realizable in real materials.\cite{kitaev2003fault,jiang2012spin,schroeter2007spin} Even with such interactions, significant debate remains, in some cases, over the nature of the ground states at the model level due to the difficulty of obtaining exact results in the presence of frustrated interactions.\cite{yan2011spin} For this reason, great excitement was generated by the exactly solvable model by Kitaev for tricoordinate lattices, which exhibits a spin liquid ground state for anisotropic but realistic short range interactions:\cite{kitaev2006anyons} 
\begin{align}
\mathcal{H} = \sum_{\langle i,j\rangle} S_i^\gamma S_j^\gamma
\end{align}
where $\gamma = \{x,y,z\}$ for the three nearest neighbour bonds emerging from each lattice point. The possibility for engineering such Kitaev interactions in real materials with strong spin-orbit coupling was advanced by the seminal work of Jackeli and Khaliullin,\cite{jackeli2009mott} who provided the following criteria: heavy $d^5$ metals in an edge-sharing octahedral environment with metal-ligand-metal bond angles close to $90^\circ$. To date, three candidate materials on the honeycomb lattice have emerged that apparently satisfy these requirements: $\alpha$-RuCl$_3$, Na$_2$IrO$_3$, and the $\alpha$-phase of Li$_2$IrO$_3$.\cite{singh2012relevance,Singh:2010gg,Plumb:2014hh} Other systems with 3D lattices, $\beta$- and $\gamma$-Li$_2$IrO$_3$ have also been discovered.\cite{modic2014realization, takayama2015hyperhoneycomb} However, all such materials exhibit an ordered ground state,\cite{choi2012spin,Johnson:2015vb,biffin2014unconventional,biffin2014noncoplanar} rather than the desirable spin-liquid, raising the question of how to engineer the Kitaev state in real systems. The above three honeycomb systems serve as a focus of the present study.

The materials of interest adopt a monoclinic $C2/m$ structure,\cite{Johnson:2015vb,choi2012spin,singh2012relevance} which is of sufficiently low symmetry that with the inclusion of SOC, there are many independent and potentially relevant terms entering the effective magnetic Hamiltonian. Up to third neighbour interactions, the magnetic Hamiltonian is characterized, in principle, by 36 symmetry inequivalent parameters. Due to this large number, it is nearly impossible to extract all such parameters from experimental data, emphasizing the need for insights from {\it ab-initio} methods. Model calculations focusing on a selection of these interactions have predicted phase diagrams hosting a rich variety of classical broken symmetry and QSL states.\cite{rau2014generic,Rau:2014wc,Katukuri:2014iq,Nishimoto:2014vw,sizyuk2014importance,PhysRevB.92.024413} In each case, the specific ground state is selected by a competition between the various interactions, so that no terms can be neglected {\it a priori}. In the absence of complete estimates for all parameters, various groups have put forward simplified models to explain the observed orders that emphasize only a selection of terms,\cite{Kimchi:2015ky,rau2014generic,kimchi2011kitaev,Reuther:2014ku,Katukuri:2014iq} although the connection between such models and the real materials remains an open question. To date, initial estimates of the magnetic interactions have largely employed low order perturbative expansions around the strong coupling $U\gg t$ limit.\cite{sizyuk2014importance,Yamaji:2014be,rau2014generic,jackeli2009mott} However, an alternative description of the electronic structures was recently put forward by some of the authors
in terms of nearly itinerant quasi-molecular orbitals delocalized across the six-site plaquettes of the honeycomb lattice.\cite{Foyevtsova:2013uo,mazin20122,mazin2013origin,Yingli2015} This latter approach emphasized the proximity of the honeycomb materials to itinerancy, a scenario that would imply significant long-range interactions and poor convergence of perturbation theory in local variables. Indeed, the Coulomb repulsion is relatively weak at the heavy Ru and Ir centres in these materials, implying the $U\gg t$ limit may not be satisfied in practice.\cite{kim2015insulating} 

The purpose of this paper is to review and refine the current understanding of interactions in the known materials using both perturbative and nonperturbative methods, and to critically evaluate the potential for engineering the Kitaev spin liquid in real materials. 
The paper is organized as follows. In Sec. \ref{sec-2} we discuss the symmetries, and material parameters relevant to the $C2/m$ structures of $\alpha$-RuCl$_3$, $\alpha$-Li$_2$IrO$_3$, and Na$_2$IrO$_3$. In Sec. \ref{sec-3}, we consider the magnetic interactions that emerge at strong coupling; Sec. \ref{sec-3}A introduces revised perturbative expressions {\it exact} in $U,J_{\rm H}$, and $\lambda$, while in Sec. \ref{sec-3}B we describe the poor convergence of the perturbative method for the long-range second and third neighbour interactions. Therefore, in Sec. \ref{sec-4}, we employ both perturbative and nonperturbative exact diagonalization techniques to estimate all magnetic interactions up to third neighbour in the three materials. The results suggest for all three materials $\alpha$-RuCl$_3$, $\alpha$-Li$_2$IrO$_3$, and Na$_2$IrO$_3$, that the classical order is selected by long-range second and third neighbour interactions. In Sec. \ref{sec-5}, we discuss the realistic range of magnetic interactions that may be realized in real materials, which allows for critical discussion about the potential for reaching a Kitaev spin liquid ground state. Finally, in Sec. \ref{sec-6}, we summarize the important conclusions.

\section{Electronic Structures}
\label{sec-2}
\subsection{General Hamiltonian}
%edit
For $d^5$ filling in an octahedral environment, one hole occupies the three $t_{2g}$ orbitals, on average, per site. We therefore consider a total Hamiltonian for the $t_{2g}$ orbitals in a nearly octahedral crystal field:
\begin{align}
\mathcal{H}_{tot} = \mathcal{H}_{\rm hop}+\mathcal{H}_{\rm CF}+\mathcal{H}_{\rm SO} + \mathcal{H}_{U} \label{eqn-1}
\end{align}
which is the sum of, respectively, the kinetic hopping term, crystal field splitting, spin-orbit coupling, and Coulomb interactions. The Coulomb terms are given by:
\begin{align}
\mathcal{H}_{U}& \ = U \sum_{i,a} n_{a,\uparrow}n_{i,a,\downarrow} + (U^\prime - J_{\rm H})\sum_{i,a< b, \sigma}n_{i,a,\sigma}n_{i,b,\sigma} \nonumber \\
 &+ U^\prime\sum_{i,a\neq b}n_{i,a,\uparrow}n_{i,b,\downarrow} - J_{\rm H} \sum_{i,a\neq b} c_{i,a\uparrow}^\dagger c_{i,a\downarrow} c_{i,b\downarrow}^\dagger c_{i,b\uparrow}\nonumber \\ & + J_{\rm H} \sum_{i,a\neq b}c_{i,a\uparrow}^\dagger c_{i,a\downarrow}^\dagger c_{i,b\downarrow}c_{i,b\uparrow} 
\end{align}
where $c_{i,a}^\dagger$ creates a hole in orbital $a\in\{xy,yz,xz\}$ at site $i$; $J_{\rm H}$ gives the strength of Hund's coupling, $U$ is the {\it intra}orbital Coulomb repulsion, and $U^\prime=U-2J_{\rm H}$ is the {\it inter}orbital repulsion.  For $5d$ Ir$^{4+}$, we take $U=1.7$ eV, $J_{\rm H}=0.3$ eV,\cite{Yamaji:2014be} while for $4d$ Ru$^{3+}$ we employ $U=3.0$ eV, $J_{\rm H}=0.6$ eV.\cite{Kim:2015iq} The one particle terms are most conveniently written in terms of:
\begin{align}
\vec{\mathbf{c}}_i^\dagger = \left(c^\dagger_{i,yz,\uparrow} \  c^\dagger_{i,yz,\downarrow} \ c^\dagger_{i,xz,\uparrow} \  c^\dagger_{i,xz,\downarrow} \ c^\dagger_{i,xy,\uparrow} \  c^\dagger_{i,xy,\downarrow}\right) 
\end{align}
Spin-orbit coupling (SOC) is described by:
\begin{align}
\mathcal{H}_{\rm SO}=\frac{\lambda}{2} \sum_i \vec{\mathbf{c}}_{i}^\dagger\left(\begin{array}{ccc} 0 & -i \sigma_z & i \sigma_y \\ i \sigma_z & 0 & -i\sigma_x \\ -i \sigma_y & i\sigma_x & 0\end{array} \right)\vec{\mathbf{c}}_i
\end{align}
where $\sigma_\mu$, $\mu=\{x,y,z\}$ are Pauli matrices. The crystal-field Hamiltonian is given by:
\begin{align}
\mathcal{H}_{\rm CF}= - \sum_i \vec{\mathbf{c}}_{i}^\dagger\left\{\mathbf{E}_i\otimes \mathbb{I}_{2\times 2}\right\}\vec{\mathbf{c}}_i
\end{align}
where $\mathbb{I}_{2\times 2}$ is the $2 \times 2$ identity matrix; in the $C2/m$ space group, the crystal field tensor $\mathbf{E}_i$ is constrained by local 2-fold symmetry at each metal site to be:
\begin{align}
\mathbf{E}_i = \left(\begin{array}{ccc} 0&\Delta_{1}&\Delta_{2} \\ \Delta_{1}&0&\Delta_{2} \\ \Delta_{2} & \Delta_{2} & \Delta_{3}\end{array} \right)
\end{align}
 The hopping Hamiltonian is most generally written:
\begin{align}
\mathcal{H}_{\rm hop}= - \sum_{ij} \vec{\mathbf{c}}_{i}^\dagger \ \left\{\mathbf{T}_{ij} \otimes \mathbb{I}_{2\times 2}\right\}\ \vec{\mathbf{c}}_j
\end{align}
with the hopping matrices $\mathbf{T}_{ij}$ defined for each bond connecting sites $i,j$. 

The effects of spin-orbit coupling $\mathcal{H}_{\rm SO}$ on the electronic structure of the $d^5$ honeycomb materials has been discussed in detail in many previous works;\cite{Foyevtsova:2013uo,sizyuk2014importance,Kim:2015iq,jackeli2009mott,Bhattacharjee:2012hk} here we discuss briefly the relevant details. The magnitude of $\mathcal{H}_{\rm SO}$ is given by the spin-orbit constant $\lambda$, for which we take $\lambda_\text{Ir} = 0.4$ eV,\cite{Kim:2014kn} and $\lambda_\text{Ru} = 0.15$ eV.\cite{Banerjee:2015wi} In the limit $\lambda\gg (\Delta,|\mathbf{T}_{ij}|)$, the local $t_{2g}$ states are split into $j_{\text{eff}}=1/2$ and $j_{\text{eff}}=3/2$ spin-orbital combinations $|j,m_j\rangle$. At each site, a single hole occupies the $j_{\text{eff}}=1/2$ states, with energy $E=+\lambda$:
\begin{align}
\left|  \frac{1}{2},\frac{1}{2}\right\rangle  &  = \frac{1}{\sqrt{3}}\left(-\left|
d_{xy}\uparrow\right\rangle -i\left|  d_{xz}\downarrow
\right\rangle -\left|  d_{yz}\downarrow\right\rangle\right) \\
\left|  \frac{1}{2},-\frac{1}{2}\right\rangle  &  = \frac{1}{\sqrt{3}}\left( \left|  d_{xy}\downarrow\right\rangle +i\left|
d_{xz}\uparrow\right\rangle -\left|  d_{yz}\uparrow
\right\rangle\right)
\end{align}
while the $j_{\text{eff}}=3/2$ states, with energy $E=-\lambda/2$ are unoccupied (by holes):
\begin{align}
\left|  \frac{3}{2},\frac{3}{2}\right\rangle  &  = \frac{1}{\sqrt{2}}\left(-i\left|
d_{xz}\uparrow\right\rangle -\left|  d_{yz}\uparrow
\right\rangle\right) \\
\left|  \frac{3}{2},\frac{1}{2}\right\rangle  &  =\frac{1}{\sqrt{6}}\left( 2\left|
d_{xy}\uparrow\right\rangle -i\left|  d_{xz}\downarrow
\right\rangle -\left|  d_{yz}\downarrow\right\rangle \right) \\
\left|  \frac{3}{2},-\frac{1}{2}\right\rangle  &  =\frac{1}{\sqrt{6}}\left( 2\left|  d_{xy}\downarrow\right\rangle-i\left|
d_{xz}\uparrow\right\rangle +\left|  d_{yz}\uparrow
\right\rangle\right) \\
\left|  \frac{3}{2},-\frac{3}{2}\right\rangle  &  =\frac{1}{\sqrt{2}}\left(-i\left|
d_{xz}\downarrow\right\rangle +\left|  d_{yz}\downarrow
\right\rangle\right)
\end{align}
For $U\gg t$, the local magnetic degrees of freedom are essentially $j_{\text{eff}}=1/2$ doublets. That is, while finite crystal field splitting and hopping tend to mix these states with the excited $j_{\text{eff}}=3/2$ states, for $\lambda \gg \Delta, t^2/U$ the low-energy states are nonetheless adiabatically connected to the $j_{\text{eff}}=1/2$ doublets.

\subsection{Hopping Integrals\label{sec_hop}}
In order to estimate the magnitude of crystal field $\mathbf{E}_i$ and hopping $\mathbf{T}_{ij}$ for the candidate honeycomb materials, we have performed {\it ab-initio} density functional theory (DFT) calculations with the linearized augmented plane wave (LAPW) method\cite{Blaha2001} for Na$_2$IrO$_3$,\cite{choi2012spin} $\alpha$-Li$_2$IrO$_3$\cite{Gretarsson2013,Manni2014} and $\alpha$-RuCl$_3$.\cite{Johnson:2015vb} The Perdew-Burke-Ernzerhof generalized gradient approximation \cite{Perdew1996} was used, with a mesh of 12 $\times$ 12 $\times$ 12 {\bf k} points in the first Brillouin zone and RK$_{\rm max}$ was set to 8. The Ir $5d$ and Ru $4d$ $t_{2g}$ hopping parameters were obtained through the Wannier function projection formalism proposed in Refs. \onlinecite{Aichorn2009,Ferber2014,Foyevtsova:2013uo}. Importantly, relativistic effects were not included in these calculations in order to not `double-count' SOC. Computed crystal-field splitting and nearest neighbour hopping integrals are given in Table \ref{tab-hops}. Further details, and full hopping integrals up to third nearest neighbour are given in Appendix \ref{app_hop}. Computed hopping integrals for Na$_2$IrO$_3$ are in excellent agreement, but differ slightly from those of Ref. \onlinecite{Foyevtsova:2013uo} and \onlinecite{mazin2013origin}, due to a finer $k$-point mesh employed in this work. 

\begin{figure}[t]
\includegraphics[width=\linewidth]{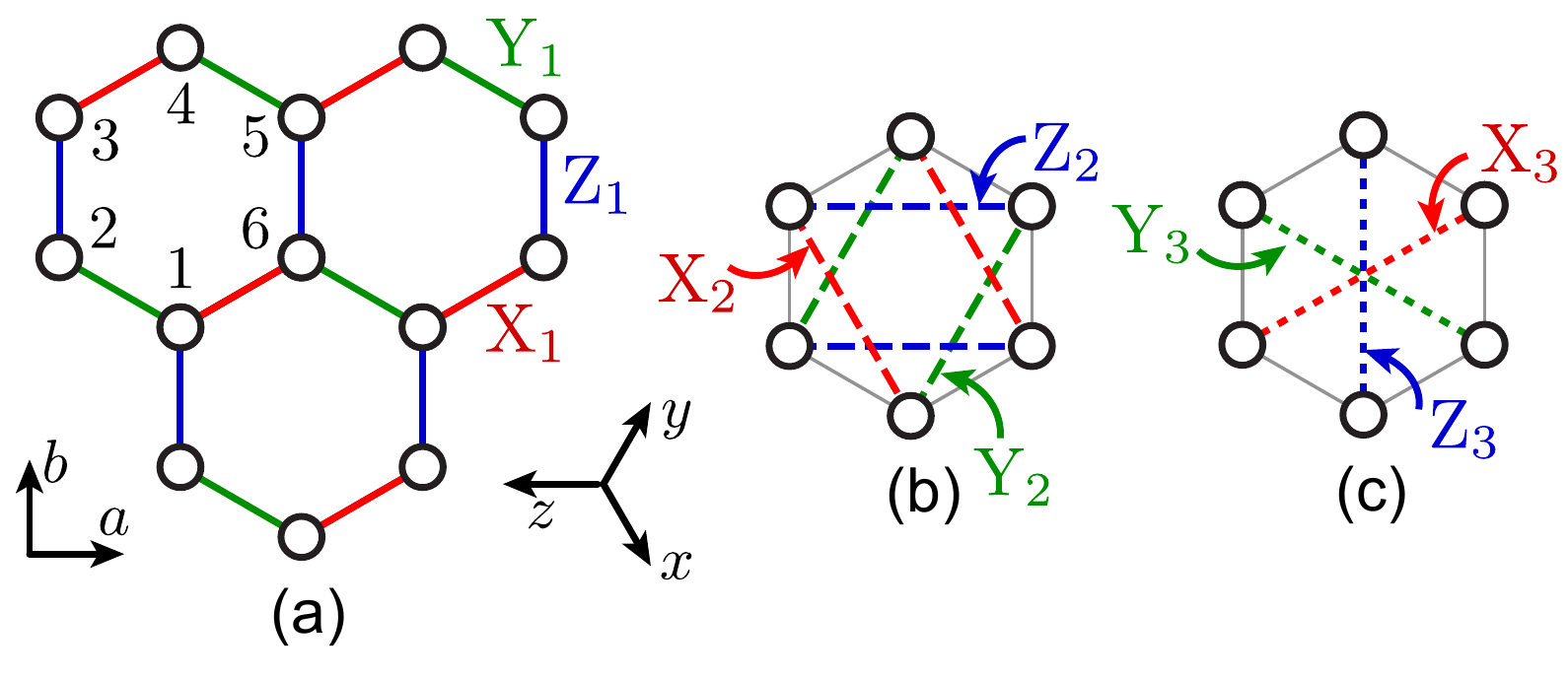}
\caption{\label{fig-labels} Cartoon of the honeycomb structure showing bond labels for (a) first neighbours, (b) second neighbours, and (c) third neighbours. Sites within a given hexagon are labelled $1-6$; $a,b$ refer to the crystallographic axes, while $x,y,z$ are the cubic axes of the local metal octahedra.}
\end{figure}

\begin{figure}[t]
\includegraphics[width=\linewidth]{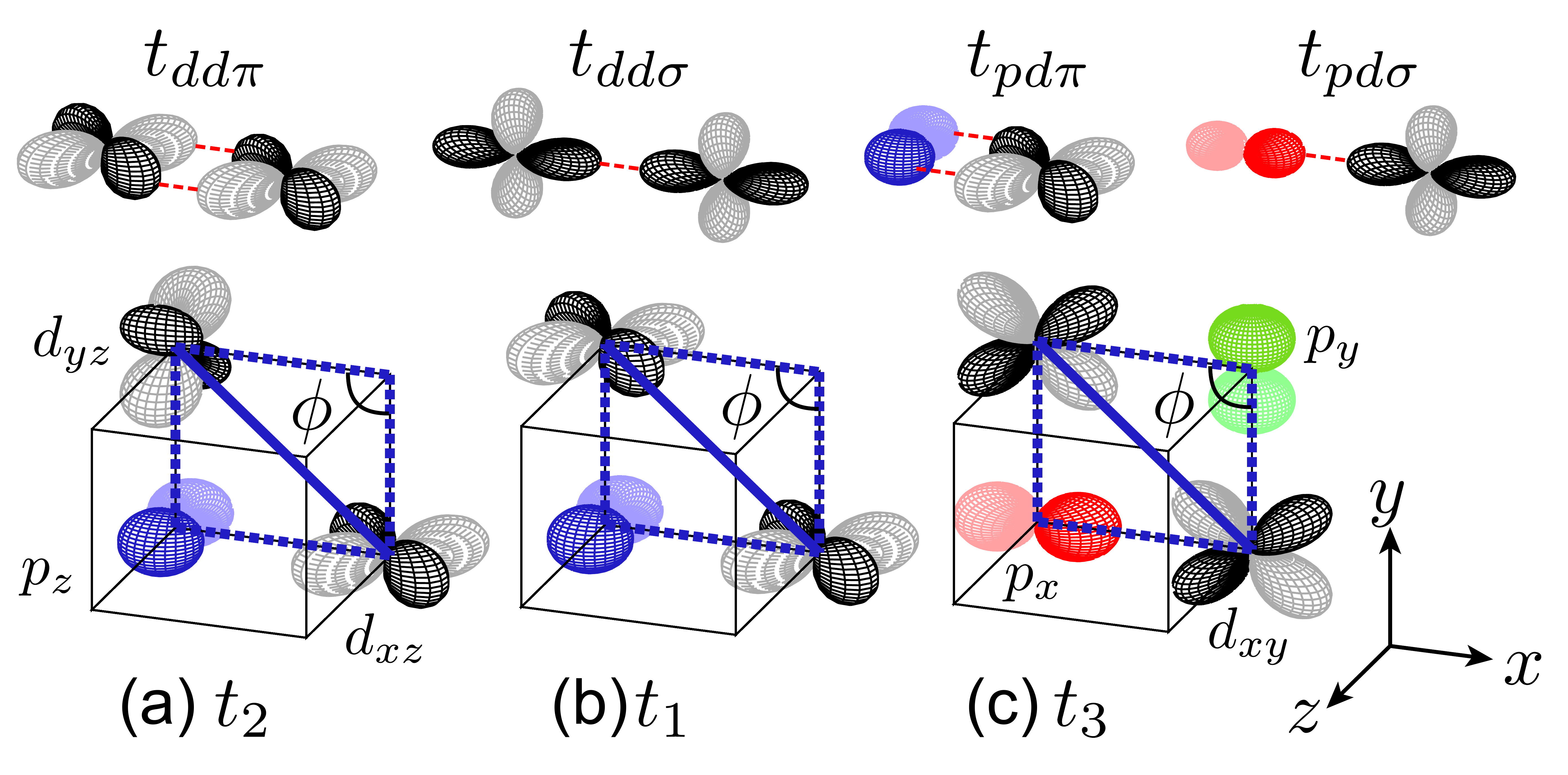}
\caption{\label{fig-SK} Geometry of nearest neighbour hopping integrals (a) $t_2$, (b) $t_1$, and (c) $t_3$ for the Z$_1$ bond, showing ligand mediated and direct hopping processes. These can be decomposed in terms of Slater-Koster hopping integrals $t_{dd\pi},t_{dd\sigma},t_{pd\pi},t_{pd\sigma}$ (top) as a function of metal-ligand-metal bond angle $\phi$, as in eqn's (\ref{eqn-hop1})-(\ref{eqn-hop2}). }
\end{figure}

The $C2/m$ space group provides two types of symmetry inequivalent nearest neighbour bonds: the so-called Z$_1$-bonds, parallel to the crystallographic $b$-axis, are of local $C_{2h}$ symmetry; we parameterize the hopping integrals for this case via $t_{1-4}$ of Ref. \onlinecite{rau2014generic}:
\begin{align}
\mathbf{T}_1^{Z}=\left(\begin{array}{ccc}t_1 & t_2 & t_4 \\ t_2 & t_1 & t_4 \\ t_4 & t_4 & t_3 \end{array} \right) 
\end{align}
where $\mathbf{T}_n^\gamma$ refers to the hopping matrix for the $n$th neighbour bond of $\gamma\in \{\text{X,Y,Z}\}$ symmetry as shown in
Fig. \ref{fig-labels} (see also Fig.~\ref{fig-SK}). The so-called X$_1$- and Y$_1$-bonds, both falling in the $ab$-plane, are of lower local $C_i$ symmetry, and therefore require additional hopping parameters:
\begin{align}
\mathbf{T}_1^{X}=\left(\begin{array}{ccc}t_3^\prime&t_{4a}^\prime&t_{4b}^\prime\\t_{4a}^\prime&t_{1a}^\prime&t_2^\prime\\t_{4b}^\prime&t_2^\prime&t_{1b}^\prime\end{array} \right) \ , \ \mathbf{T}_1^{Y}=\left(\begin{array}{ccc}t_{1a}^\prime&t_{4a}^\prime&t_2^\prime\\t_{4a}^\prime&t_3^\prime&t_{4b}^\prime\\t_2^\prime&t_{4b}^\prime&t_{1b}^\prime\end{array} \right) 
\end{align}
In the absence of significant distortions, one should expect nearly $C_3$ symmetry of the hopping such that $t_n\approx t_{na}^\prime \approx t_{nb}^\prime$; in general, these relationships do not hold exactly, and some bond-dependence results. 
\begin{table}[t]
\caption {\label{tab-hops}Parameters for crystal field splitting and nearest neighbour hopping (meV) for experimental $C2/m$ structures of Na$_2$IrO$_3$,\cite{choi2012spin} $\alpha$-Li$_2$IrO$_3$,\cite{Gretarsson2013} and $\alpha$-RuCl$_3$.\cite{Johnson:2015vb} Hopping integrals are labelled according to Ref. \onlinecite{rau2014generic}; in brackets are given the corresponding labels from Ref. \onlinecite{Foyevtsova:2013uo}.}
\begin{ruledtabular}
\begin{tabular}{cccc}
Term&Na$_2$IrO$_3$&$\alpha$-Li$_2$IrO$_3$&$\alpha$-RuCl$_3$\\
\hline
$\Delta_1$&-22.9&-37.5&-19.8 \\
$\Delta_2$&-27.6&-35.0&-17.5 \\
$\Delta_3$& -27.2&-5.5&-12.5\\
\hline
$t_1$ ($t_{\bar{1}||}$)&+33.1&+55.0&+50.9\\
$t_{1a}^\prime$ ($t_{1||}$)&+29.9&+80.2&+44.9\\
$t_{1b}^\prime$ ($t_{1||}$)&+47.6&+72.3&+45.8\\
\hline
$t_2$ ($t_{\bar{1}\text{O}}$)&+264.3&+219.0&+158.2 \\
$t_{2}^\prime$ ($t_{1\text{O}}$)&+269.3&+252.7&+162.2\\
\hline
$t_3$ ($t_{\bar{1}\sigma}$)&+26.6&-175.1&-154.0 \\
$t_{3}^\prime$ ($t_{1\sigma}$)&-19.4&-108.8&-103.1\\
\hline
$t_4$ ($t_{\bar{1}\perp}$)&-11.8&-124.5&-20.2\\
$t_{4a}^\prime$ ($t_{1\perp}$)&-21.4&-16.7&-15.1\\
$t_{4b}^\prime$ ($t_{1\perp}$)&-25.4&-1.9&-10.9
\end{tabular}
\end{ruledtabular}
\end{table}

For the experimental structures of all three studied materials, we find relatively small crystal-field splitting, such that $\lambda \gtrsim 10 \Delta_n$. For nearest neighbour bonds, the largest integrals are $t_2\approx t_2^\prime$ and $t_3\approx t_3^\prime$, arising predominantly from hopping either through the bridging ligand oxygen or halogen $p$-orbitals ($t_2$), or direct metal-metal hopping ($t_3$). The origin of these hopping terms is discussed in more detail in Sec. \ref{sec-5}. It is well known that for Na$_2$IrO$_3$, distortion of the IrO$_6$ octahedra by the large Na$^+$ ion elongates Ir-Ir distances to $3.13-3.14$ \AA, while Ir-O-Ir bond angles are as large as 100$^\circ$.\cite{choi2012spin} While this distortion apparently does not significantly enhance the crystal field terms $\Delta_n$, it does suppress direct hopping, causing $t_2$ to dominate ($|t_2/t_3| \sim 10$). The bond-anisotropy for Na$_2$IrO$_3$ is also small (i.e. $t_2\approx t_2^\prime$), which is suggestive of small bond-dependence in the magnitude of the resulting magnetic interactions. In contrast, $\alpha$-Li$_2$IrO$_3$ and $\alpha$-RuCl$_3$ display much greater direct hopping, with ($|t_2/t_3| \sim 1$). For the experimentally determined structure of $\alpha$-Li$_2$IrO$_3$, the smaller Li$^+$ ion is more easily incorporated, such that Ir-Ir distances are reduced to $2.98-2.99$ \AA, and Ir-O-Ir bond angles $\sim 94^\circ$.\cite{Gretarsson2013} For this material, we also find significant bond-anisotropy, particularly in $t_4 \gg t_{4a}^\prime,t_{4b}^\prime$; the results of this finding on the magnetic interactions is discussed in detail in section \ref{sec_Li}. For $\alpha$-RuCl$_3$, the recently revised $C2/m$ structure provides similar Ru-Cl-Ru bond angles $\sim 94^\circ$,\cite{Johnson:2015vb} which also allow for large direct hopping $t_3,t_3^\prime$.

Full details of the computed long-range second and third neighbour hopping is given in Appendix \ref{app_hop}. As shown in Fig. \ref{fig-labels}, we label long-range hopping matrices $\mathbf{T}_{n}^\gamma$ by the character of the intervening nearest neighbour bonds. Second neighbours ($n=2$) share a bond label ($\gamma$) with the perpendicular first neighbour bond, such that those joined by intervening nearest neighbour X$_1$ and Y$_1$ bonds, are labelled Z$_2$. Similarly, those linked by (Y$_1$, Z$_1$) bonds and (Z$_1$, X$_1$) are labelled X$_2$ and Y$_2$, respectively.  The largest second neighbour hopping integrals computed are $\sim 50-70$ meV, between $d$ orbitals sharing a label with the bond-type, e.g. $d_{xz}\rightarrow d_{xy}$ for the X$_2$ bonds. Third neighbours are labelled via the parallel first neighbour bond. The third neighbour bonds have the same symmetry as the corresponding first neighbour bond. The largest hopping integrals were found to be $\sim 30-40$ meV, between $d$-orbitals not sharing a label with the bond-type, e.g. $d_{yz}\rightarrow d_{yz}$ for the X$_3$ bond. In both cases, these largest second and third neighbour hopping integrals arise from M-L-L-M paths (M = metal, L = ligand) that are promoted by the close L-L contacts at or within the van der Waals radii.

\section{Magnetic Interactions}
\label{sec-3}

\subsection{General Form}

\begin{figure*}[t]
\includegraphics[width=0.85\linewidth]{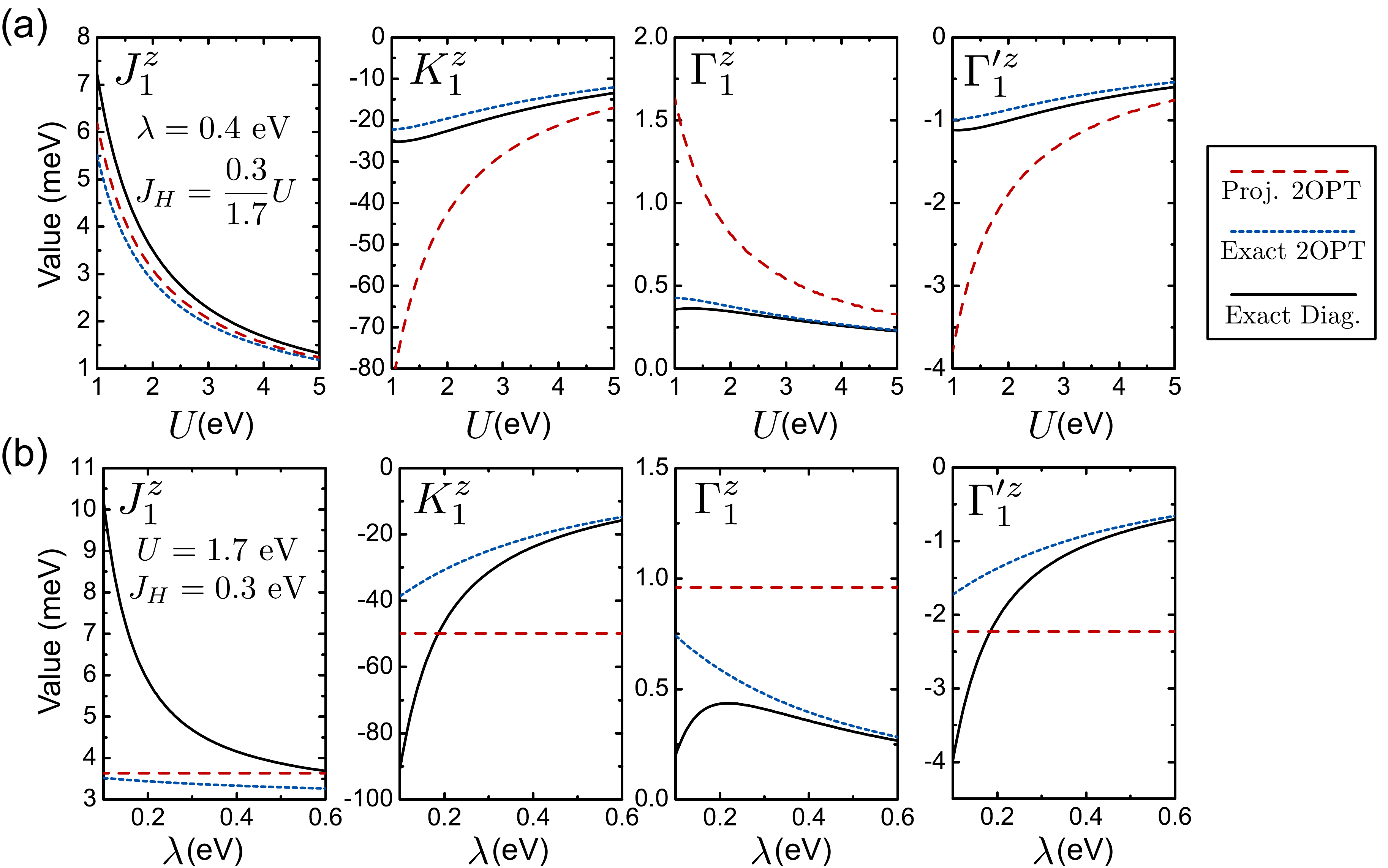}
\caption{\label{fig-UL} Nearest neighbour interactions for the Z$_1$-bond in Na$_2$IrO$_3$ in the absence of crystal-field splitting, employing hopping parameters described in section \ref{sec_hop}. The results of 2-site exact diagonalization (black solid line) are compared with approximate projective expressions (red dashed line, Ref. \onlinecite{rau2014generic}) and second order perturbation theory exact in $\lambda, U, J_{\rm H}$ (blue dotted line). (a) $U$-dependence, with constant $J_{\rm H}/U$ ratio of $0.3/1.7$, and $\lambda=0.4$ eV. (b) $\lambda$-dependence, with $J_{\rm H} = 0.3$ eV, and $U = 1.7$ eV. }
\end{figure*}

In the limit $U\gg t$, holes occupying the $j_{\text{eff}}=1/2$ states are nearly localized to their parent metal sites, and the low energy degrees of freedom are pseudo-spin $1/2$ variables $\mathbf{S}_i$ adiabatically connected to the $j_\text{eff}=1/2$ states discussed in the previous section.\cite{sizyuk2014importance,Rau:2014wc,Yamaji:2014be,Bhattacharjee:2012hk} In this case, the relevant Hamiltonian can be written:
\begin{align}
\mathcal{H}_{\rm spin} = \sum_{\langle ij\rangle} \mathbf{S}_i \cdot \mathbf{J}_{ij} \cdot \mathbf{S}_j+\mathcal{O}(\mathbf{S}^4) \label{eqn-10}
\end{align}
where $\langle ij\rangle$ denotes a sum over all pairs of sites. For intermediate values of $U/t$, $\mathcal{H}_{\rm spin}$ remains valid at low energies, despite relative itinerancy of the holes, but the corresponding magnetic interactions become increasingly nonlocal. In this work, we consider up to third neighbour ($n=3$) interactions. In the absence of any relevant symmetries, the interaction matrices $\mathbf{J}_{ij} = \mathbf{J}_{ij,s} + \mathbf{J}_{ij,a}$ are conventionally parameterized in terms of symmetric ($\mathbf{J}_{ij,s}$) and antisymmetric ($\mathbf{J}_{ij,a}$) components:
\begin{align}
\mathbf{J}_{ij,s}=& \ \left(\begin{array}{ccc} J_{ij}+\Gamma_{ij}^{aa} &\Gamma_{ij}^{ab} & \Gamma_{ij}^{ac} \\ \Gamma_{ij}^{ab} & J_{ij}+\Gamma_{ij}^{bb} & \Gamma_{ij}^{bc} \\ \Gamma_{ij}^{ac}& \Gamma_{ij}^{bc} & J_{ij}+\Gamma_{ij}^{cc}\end{array}\right) \\
\mathbf{J}_{ij,a}=& \ \left(\begin{array}{ccc} 0 &D_{ij}^c& -D_{ij}^b \\ -D_{ij}^c& 0 & D_{ij}^a\\ D_{ij}^b& -D_{ij}^a & 0\end{array}\right)
\end{align}
which corresponds to the Hamiltonian:
\begin{align}
\mathcal{H} = \sum_{\langle ij\rangle } J_{ij} \  \mathbf{S}_i \cdot \mathbf{S}_j + \mathbf{D}_{ij} \cdot \mathbf{S}_i \times \mathbf{S}_j + \mathbf{S}_i \cdot \mathbf{\Gamma}_{ij} \cdot\mathbf{S}_j
\end{align}
where $J_{ij}$ is the scalar Heisenberg coupling, $\mathbf{D}_{ij}=(D_{ij}^a,D_{ij}^b,D_{ij}^c)$ is the Dzyaloshinskii-Moriya (DM) vector, and the traceless tensor $\mathbf{\Gamma}_{ij}$ characterizes the pseudo-dipolar interaction. In Kitaev's original honeycomb model, the interactions are bond-dependent and described by:
\begin{align}
\mathcal{H}_\text{Kitaev} = \sum_{\text{1 nn}} S_i^\gamma S_j^\gamma
\end{align}
where $\gamma\in\{x,y,z\}$ for the X$_1$, Y$_1$, and Z$_1$ bonds, respectively. 
In order to emphasize this interaction, it is convenient to rewrite the symmetric part of $\mathbf{J}_{ij}$ in a $\gamma$-dependent form. For the Z$_n$-bonds, symmetry allows:\cite{rau2014generic}
\begin{align}\label{eqn-a1}
\mathbf{J}_{n,s}^Z=\left(\begin{array}{ccc} J_{n}^{z} & \Gamma_n^z & \Gamma^{\prime z}_n \\ \Gamma_n^z & J_n^z & \Gamma^{\prime z}_n \\ \Gamma^{\prime z}_n & \Gamma^{\prime z}_n & J_n^z+K_{n}^{z}\end{array}\right)
\end{align}
in terms of the isotropic exchange $J_n$, Kitaev coupling $K_n$, and off-diagonal anisotropic terms $\Gamma_n,\Gamma^\prime_n$. The superscript $z$ denotes values appropriate for the Z$_n$ bonds. For the lower symmetry X$_n$- and Y$_n$-bonds, six unique parameters are allowed by symmetry:
\begin{align}
\mathbf{J}_{n,s}^X=\left(\begin{array}{ccc} J_n^{xy}+K_1^{xy} & \Gamma_n^{\prime xy}+\zeta_n & \Gamma_n^{\prime xy}-\zeta_n \\ \Gamma_n^{\prime xy}+\zeta_n & J_n^{xy}+\xi_n & \Gamma_n^{xy} \\ \Gamma_n^{\prime xy}-\zeta_n& \Gamma_n^{xy} & J_n^{xy}-\xi_n\end{array}\right)\\
\mathbf{J}_{n,s}^Y=\left(\begin{array}{ccc} J_n^{xy} +\xi_n  & \Gamma_n^{\prime xy}+\zeta_n & \Gamma_n^{xy}\\ \Gamma_n^{\prime xy}+\zeta_n & J_n^{xy}+K_n^{xy}&  \Gamma_n^{\prime xy}-\zeta_n \\ \Gamma_n^{xy}&  \Gamma_n^{\prime xy}-\zeta_n & J_n^{xy}-\xi_n\end{array}\right) \label{eqn-a3}
\end{align}
Previous studies have typically assumed C3 symmetry of the interactions, such that $\xi_n,\zeta_n \sim 0$, $J_n^z \approx J_n^{xy}$, $K_n^z \approx K_n^{xy}$, $\Gamma_n^z\approx\Gamma_n^{xy}$, and $\Gamma_n^{\prime z} \approx\Gamma_n^{\prime xy}$. In the $C2/m$ space group, inversion symmetry requires that $\mathbf{D}_{ij}$, and therefore $\mathbf{J}_{ij,a}$ vanishes for all first and third neighbour bonds, such that $\mathbf{J}_n^\gamma = \mathbf{J}_{n,s}^\gamma$ for $n=1,3$. For second neighbours, symmetry allows the DM-interaction; in the explicit calculations below, we therefore present $\mathbf{D}_{ij}$ for all second neighbour pairs.

\subsection{Nearest Neighbours for $\Delta_n=0$\label{sec_2OPT}}

For $\Delta_n=0$, various analytical expressions for Z$_1$-bond interactions $J_1^z,K_1^z,\Gamma_1^z,\Gamma^{\prime z}_1$ obtained from perturbation theory at $\mathcal{O}(t^2)$, have appeared recently in the literature.\cite{rau2014generic,jackeli2009mott} The most widely used are based on projection of the $\lambda = 0$ Kanamori-type Hamiltonian for the $t_{2g}$ orbitals onto the relativistic $j_{\text{eff}} = 1/2$ basis. This procedure becomes exact only in the unphysical limit $U\gg\lambda\gg t$ which is not satisfied generally in real materials. In order to improve on these results, we have computed expressions exact to {\it all} orders of $J_{\rm H}, U, \lambda$ in the absence of crystal field splitting ($\Delta_n=0$). In this section, we consider the case for the nearest neighbour Z$_1$-bonds for ideal octahedral bond geometry with all metal-ligand-metal bond angles $90^\circ$. Expressions for general hoppings are given in Appendix \ref{app_terms}. In section IV we will generalize the results to the case $\Delta_n \neq 0$ using exact diagonalization (ED) calculations for the real materials of interest.

At $\mathcal{O}(t^2)$, magnetic interactions result from a combination of i) ``{\it intra}band'' terms ($\propto \mathbb{A}>0$) arising from virtual hopping of holes between $j_{\text{eff}} = 1/2$ states, and ii) ``{\it inter}band'' terms ($\propto \mathbb{B}>0$) arising from hopping between $j_{\text{eff}} = 1/2$ and lower-lying $j_{\text{eff}} = 3/2$ states. Both processes contribute to the isotropic exchange $J_1$, but with opposite sign:
\begin{align}
J_1^z =& \ \frac{4\mathbb{A}}{9}(2t_1+t_3)^2-\frac{8\mathbb{B}}{9}\left\{9t_4^2+2(t_1-t_3)^2\right\} \label{eqn-J1}
\end{align}
while the anisotropic terms arise only from interband processes ($\propto \mathbb{B}$):
\begin{align}
K_1^z =& \frac{8\mathbb{B}}{3}\left\{ (t_1-t_3)^2+3t_4^2-3t_2^2\right\}\label{eqn-K1} \\
\Gamma_1^z =& \ \frac{8\mathbb{B}}{3}\left\{ 3t_4^2+2t_2(t_1-t_3)\right\}\label{eqn-G1} \\
\Gamma^{\prime z}_1 =& \ \frac{8\mathbb{B}}{3}\left\{ t_4(3t_2+t_3-t_1)\right\} \label{eqn-Gp1}
\end{align}
The constants appearing in these expressions are derived from the propagator with respect to $\mathcal{H}_{U}+\mathcal{H}_{\rm SO}$ for a single hole added to the $t_{2g}$ states (see Appendix \ref{app_terms}), and can be computed exactly:\footnote{We note that the limit $U,\lambda \gg J_{\rm H}$ has already been considered in Ref. \onlinecite{rau2014generic}, although the obtained expressions were not analyzed. In this case, the constants reduce to $\mathbb{A} \approx  \left(3U+4J_{\rm H} \right)/\left(3U^2\right)$ and $\mathbb{B} \approx  \left(4J_{\rm H}\right)/\left(3(2U+3\lambda)^2\right) $.
It is easy to verify that Eqs. (\ref{eqn-J1})$-$(\ref{eqn-Gp1}) agree with those provided in the supplemental material of Ref. \onlinecite{rau2014generic}.}
\begin{align}
\mathbb{A} = & \ -\frac{1}{3}\left\{\frac{J_{\rm H} + 3(U+3\lambda)}{6J_{\rm H}^2 - U(U+3\lambda)+J_{\rm H}(U+4\lambda)} \right\}\\
\mathbb{B} = & \ \frac{4}{3}\left\{\frac{(3J_{\rm H}-U-3\lambda)}{(6J_{\rm H}-2U-3\lambda)}\eta\right\}\\
\eta= & \ \frac{J_{\rm H}}{6J_{\rm H}^2-J_{\rm H}(8U+17\lambda)+(2U+3\lambda)(U+3\lambda)} 
\end{align}
The values of these constants can be estimated for the real materials; for $5d$ Ir$^{4+}$ ions (as in $A$$_2$IrO$_3$, $A$ = Na, Li), we take $U=1.7$, $J_{\rm H}=0.3$, and $\lambda=0.4$ eV, suggesting:
\begin{align}
 \mathbb{A}_{5d}\sim 0.9 \text{ eV}^{-1} \ \ , \ \  \mathbb{B}_{5d}\sim 0.04 \text{ eV}^{-1}\end{align}
while for $4d$ Ru$^{3+}$ ions (as in $\alpha$-RuCl$_3$), we take $U=3.0$, $J_{\rm H}=0.6$, and $\lambda=0.15$ eV, suggesting:
\begin{align}
 \mathbb{A}_{4d}\sim 0.6 \text{ eV}^{-1} \ \ , \ \  \mathbb{B}_{4d}\sim 0.05 \text{ eV}^{-1}\end{align}
The second order expressions may be compared with the results of exact diagonalization (ED) of the full Hamiltonian $\mathcal{H}_{tot}$ on two sites (for $\Delta_n=0$). In the latter case, the interaction parameters $J_1^z,K_1^z,$ etc. were extracted via projection of the exact low-energy states onto the $j_{\text{eff}}=1/2$ states as described in Appendix \ref{app_ED}. We show in Fig. \ref{fig-UL} the dependence of the interactions on $\lambda$ and $U$ for Hamiltonian parameters suitable for the Z$_1$-bond of Na$_2$IrO$_3$. For the $\lambda$-dependence plots, $U = 1.7,J_{\rm H}=0.3$ eV are fixed, while $U$-dependence is considered with fixed $\lambda=0.4$ eV and $J_{\rm H}/U$ ratio. One can see that the  ``exact'' second order expressions (\ref{eqn-J1})$-$(\ref{eqn-Gp1}) agree with the ED results over a wide range of $U$-values, and break down only in the weak $\lambda$ limit. Interestingly, large $\lambda$ tends to suppress the anisotropic terms, due to enhancement of the gap between the $j_{\text{eff}}=1/2$ and $j_{\text{eff}}=3/2$ states. The close agreement between the perturbative and ED results validates both approaches. In contrast, the projective expressions of Ref. \onlinecite{rau2014generic} seem to overestimate the magnitude of the anisotropic terms over a large region of parameters, and fail to capture any $\lambda$-dependence by construction. 

In real materials, $\mathbb{A} \gg \mathbb{B}$, so that the anisotropic interactions typically represent subleading terms. For materials close to the Kitaev limit ($K_1\gg J_1$), the leading term $J_1$ must therefore be suppressed to an order of magnitude below its natural scale,\cite{jackeli2009mott} which opens the possibility that other subleading interactions such as second and third neighbour terms may also be relevant. These are considered in the next section.

% In the next section, we consider also higher order corrections, and longer-range interactions, which may be extracted from ED calculations on real materials.

\subsection{Long-Range Interactions}

\begin{figure}[t]
\includegraphics[width=0.95\linewidth]{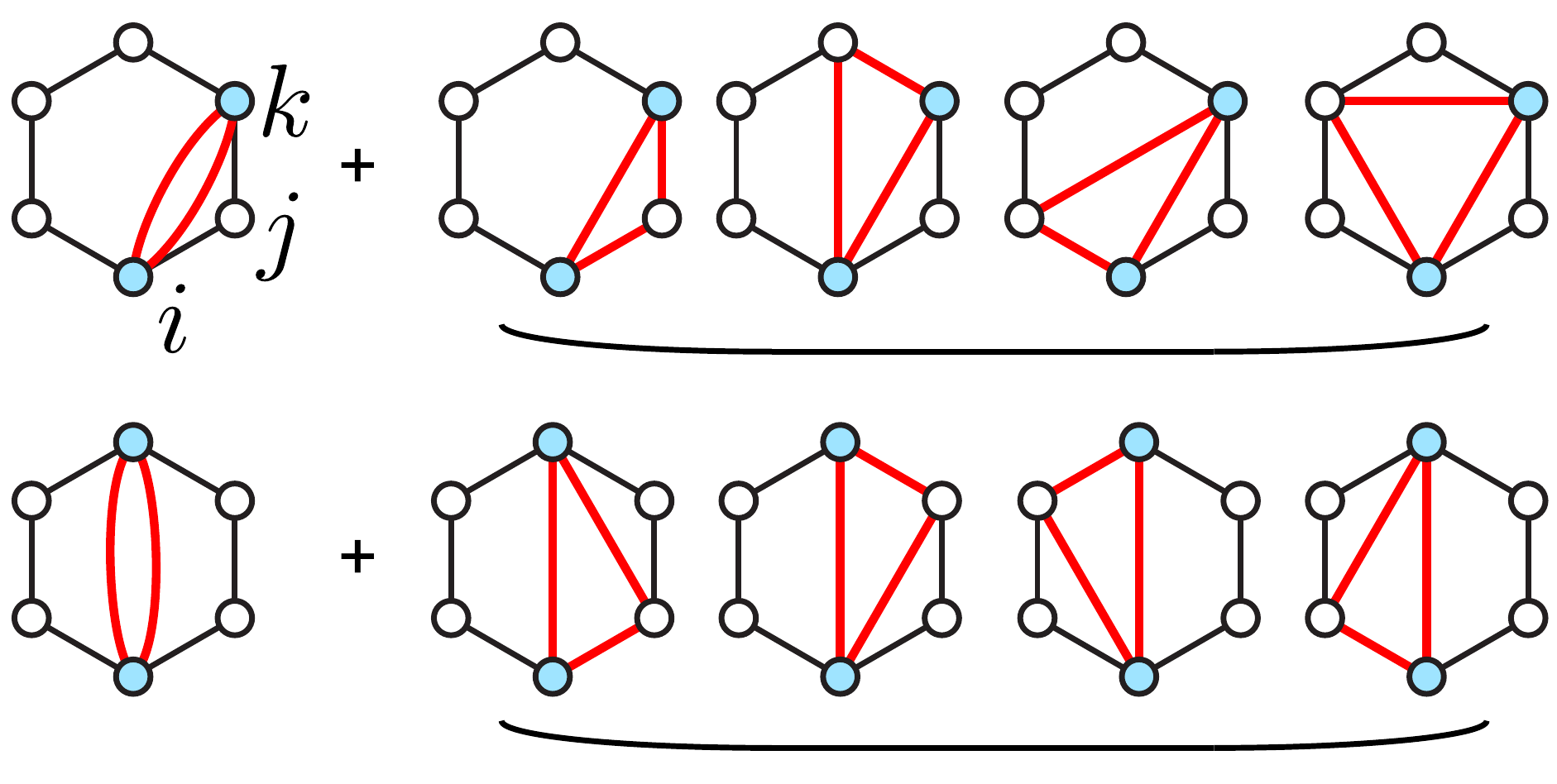}
\caption{\label{fig-highorder} Hopping paths associated with contributions $\mathcal{O}(t^2)$ and $\mathcal{O}(t^3)$ to the long-range second neighbour (top) and third neighbour (bottom) interactions. The many higher order corrections to these terms must be included to produce accurate estimates.}
\end{figure}

Various previous works have considered long-range terms for the honeycomb materials, either for interpretation of experimental data,\cite{kimchi2011kitaev, Reuther:2014ku,singh2012relevance,Nishimoto:2014vw,choi2012spin} or from an {\it ab-initio} perspective.\cite{Kim:2015iq,sizyuk2014importance,Yamaji:2014be,Plumb:2014hh} In the latter reports, such interactions where estimated only at the level of second order perturbation theory in the direct second or third neighbour hopping. Here we consider the validity of this approach. For second neighbour interactions, we consider three adjacent sites $i,j,k$. The lowest order contributions to the second neighbour interactions arise from direct hopping associated with virtual hopping processes such as $i\rightarrow k \rightarrow i$:
 \begin{align}
 J_2^{(2)} \sim  \frac{|\mathbf{T}_{ik}\mathbf{T}_{ki}|}{U_\text{eff}}
 \end{align}
where $U_\text{eff} (U,J_{\rm H},\lambda)\sim \mathbb{A}^{-1}\sim 1.0-1.5$ eV gives the rough energy cost for double occupancy of a given site. Such contributions have been previously considered in the literature. As shown in Fig. \ref{fig-highorder}, several virtual hopping paths contribute to $\mathcal{O}(t^3)$ terms, the largest of which provides: 
\begin{align}
J_2^{(3)} \sim  \frac{|\mathbf{T}_{ij}\mathbf{T}_{jk}\mathbf{T}_{ki}|}{U_\text{eff}^2}
\end{align}
This corresponds to the three site ring exchange $i\rightarrow j \rightarrow k \rightarrow i$ process. Strong convergence of the perturbation expansion would require $J_2^{(2)}\gg J_2^{(3)}$. However, for conservative estimates of $U_\text{eff} \sim 1$ eV, and $|\mathbf{T}_{ij}/|\mathbf{T}_{ik}| \sim 10$, the second order $J_2^{(2)}$ and third order $J_2^{(3)}$ contributions can be of similar magnitude. On this basis, we conclude that perturbation theory for the long-range interactions may not be strongly convergent, questioning
the reliability of previous estimates. This finding is consistent with previous suggestions that long-range interactions on the scale of $n=2,3$ would emerge naturally from a semi-itinerant picture of the holes within the hexagonal plaquettes~\cite{Foyevtsova:2013uo,mazin20122}. In order to bridge these two perspectives, we have applied nonperturbative exact diagonalization methods (Appendix \ref{app_ED}) to the real materials in the following sections, which allow for inclusion of crystal field splitting effects as well as accurate estimation of all terms up to third neighbour. 
We have also considered 4-spin and 6-spin ring-exchange interactions that similarly emerge at high orders in perturbation theory, but we find them to be negligible in the calculations below, implying sufficient convergence at third order.

\section{Application to Specific Materials}
\label{sec-4}
\subsection{Na$_2$IrO$_3$}
\subsubsection{Introduction}
A range of recent studies of Na$_2$IrO$_3$ have established this material as a $j_{\text{eff}}=1/2$ spin-orbit assisted Mott insulator with significantly anisotropic (and bond-dependent) interactions of the Kitaev type.\cite{Singh:2010gg,HwanChun:2015ed}
Magnetic order below $T_N = 14$ K has now been unambiguously established to be of collinear zig-zag type,\cite{Liu:2011ed,choi2012spin,Ye:2012dr} with moments ordered at 45$^\circ$ from the crystallographic $ab$-plane and cubic $x,y$-axes of the IrO$_6$ octahedra.\cite{HwanChun:2015ed} This type of order was initially unexpected as the pure nearest neighbour $(J_1,K_1)$ Heisenberg-Kitaev (nnHK) model assumed to be relevant yields instead stripy order in the parameter region suggested by perturbation theory (i.e. $J_1> 0$, $K_1<0$, and $|K_1|>|J_1|$). It was subsequently shown that the observed zigzag order could be stabilized by large second and third neighbour Heisenberg coupling ($J_2,J_3$),\cite{kimchi2011kitaev,singh2012relevance} particularly in combination with large $\Gamma_1,\Gamma_1^\prime$\cite{PhysRevB.92.024413} and/or second neighbour Kitaev coupling $K_2$.\cite{sizyuk2014importance} Given the large number of independent parameters, extraction from experiment has not yet been possible, although analysis of inelastic neutron scattering data suggested significant long-range terms.\cite{choi2012spin} In addition, evidence for significant antiferromagnetic terms can be taken from the large negative isotropic Weiss constant $\Theta_{iso} \sim -116$ K, while anisotropy in this term ($\Theta_{ab}>\Theta_c$) suggests some bond-anisotropy or off-diagonal $\Gamma_1,\Gamma_1^\prime$ terms, perhaps arising due to crystal field splitting.\cite{Singh:2010gg} 

From the perspective of {\it ab-initio} calculations, all published DFT studies find evidence for a large $t_2\gg t_1,t_3,t_4$ hopping integral, as in section \ref{sec_hop}. On this basis, it is generally well accepted that the dominant interaction in Na$_2$IrO$_3$ is indeed a ferromagnetic $K_1<0$ term, with a subdominant antiferromagnetic $J_1>0$ as originally proposed. However, a clear picture of all relevant interactions, and their relationship to zigzag order, is currently under debate. Katukuri {\it et al.}~\cite{Katukuri:2014iq} 
 employed MRCI (MultiReference Configuration Interaction) state energies for Ir dimers, to parameterize a simplified nearest neighbour $(J_1,K_1,\Gamma_1)$ model.\cite{Katukuri:2014iq}  These authors emphasized a small anisotropy between the X$_1$,Y$_1$- and Z$_1$-bonds. For the latter, they suggested $J_1= +5.0, \ K_1 = -20.5, \ \Gamma_1 = +0.5$ meV, in agreement with initial expectations.
Two additional studies have subsequently appeared, employing numerical 2$^{nd}$ order perturbation theory (N2OPT) in terms of {\it ab-initio} derived hopping and crystal field parameters. The estimates of Ref. \onlinecite{sizyuk2014importance} employed a selection of hopping integrals of Ref. \onlinecite{Foyevtsova:2013uo} and found the largest nearest neighbour terms to be $J_1 = +5.8$, $K_1= -14.8$ meV consistent with the MRCI results. Beyond nearest neighbours, the authors also suggested the possibility of large $J_2 = -4.4$, $K_2 = +7.9$ meV terms, a possibility we revisit here with a more complete treatment. In contrast with both these studies, a very large $|\Gamma_1^{\prime xy}+\zeta_1|$ term $> 8$ meV for the X$_1$,Y$_1$-bonds was found by Yamaji {\it et al.},\cite{Yamaji:2014be} and argued to be responsible for the observed zigzag order. However,  this result
may suffer from ``double-counting'' SOC since  both spin-dependent
 hopping integrals from relativistic DFT calculations {\it and} 
an on-site $\lambda \mathbf{L}\cdot\mathbf{S}$ term are included
in the perturbation theory, which may explain the differences between the computed values of this work. Moreover, the interactions computed in Ref. \onlinecite{Yamaji:2014be} predict a ferromagnetic Weiss constant $\Theta_{iso}>0$ in contradiction with experiment. For these reasons, it is of significant value to reanalyze the magnetic interactions in Na$_2$IrO$_3$ using nonperturbative methods in order to establish more accurate estimates of the magnetic exchange parameters.

\subsubsection{Calculations and Discussion}

\begin{table}[t]
\caption{ \label{table_NaIr1} Nearest neighbour magnetic interactions in meV for Na$_2$IrO$_3$ obtained from various methods employing $U$ = 1.7 eV, $J_{\rm H}$ = 0.3 eV, $\lambda$ = 0.4 eV. For CFS (Crystal Field Splitting) = ``no'', $\Delta_n = 0$. The most accurate method theoretically is 6-site ED, highlighted in bold.}
\begin{ruledtabular}
\begin{tabular}{cccccc}
\multicolumn{6}{c}{Z$_1$-bonds:}\\
\hline \\[-2.0ex]
Method&CFS&$J_1$&$K_1$&$\Gamma_1$&$\Gamma_1^\prime$\\ 
\hline
%Proj. 2OPT&No&+3.8&-49.8&+1.1&-2.2&0.87& \\
exact 2OPT&no&+3.2&-20.5&+0.4&-0.9\\
ED (2-site)&no&+4.2&-23.7&+0.3&-1.0\\
ED (2-site)&full&+1.8&-25.5&-0.4&-2.8\\
{\bf ED (6-site)}&\bf full&\bf +1.6&\bf -17.9&\bf -0.1&\bf -1.8\\[1.0ex]
\hline \\[-2.0ex]
\multicolumn{6}{c}{Literature Values}\\
\hline \\[-2.0ex]
2-site MRCI\cite{Katukuri:2014iq}&approx.&+5.0&-20.5&  +0.5&$-$  \\
N2OPT\cite{Yamaji:2014be}&full&+4.4&-35.1&-0.4&+1.1\\[1.0ex]
\end{tabular}
\end{ruledtabular}
\vspace{3mm}
\begin{ruledtabular}
\begin{tabular}{cccccccc}
\multicolumn{8}{c}{X$_1$,Y$_1$-bonds:}\\
\hline \\[-2.0ex]
Method&CFS&$J_1^{xy}$&$K_1^{xy}$&$\xi_1^{xy}$ & $\Gamma_1^{xy}$&$\Gamma_1^{\prime xy}$&$\zeta_1^{xy}$\\ 
\hline \\[-2.0ex]
exact 2OPT&no  &+0.9 &-20.9 &  -0.1&+3.3 &-1.7 & -0.1\\
ED(2-site)&no  &+1.7 &-24.1& -0.2 &+3.0 &-2.0 & -0.1 \\
ED(2-site)&full&0.0  &-23.3& +0.1 &+2.0 &-3.4 & +0.1 \\ 
\bf ED(6-site)&\bf full&\bf -0.1&\bf -16.2&\bf -0.1 &\bf+2.1&\bf -2.3&\bf +0.1\\[1.0ex] 
\hline \\[-2.0ex]
\multicolumn{8}{c}{Literature Values}\\
\hline \\[-2.0ex]
2-site MRCI\cite{Katukuri:2014iq}&approx.&+1.5&-15.2&$-$&   +1.2     &$-$&$-$ \\
N2OPT\cite{Yamaji:2014be}&full&+2.6&-27.9&+0.6&+1.8&-5.8&+2.7
\end{tabular}
\end{ruledtabular}
\end{table}

\begin{figure}[t]
\includegraphics[width=0.7\linewidth]{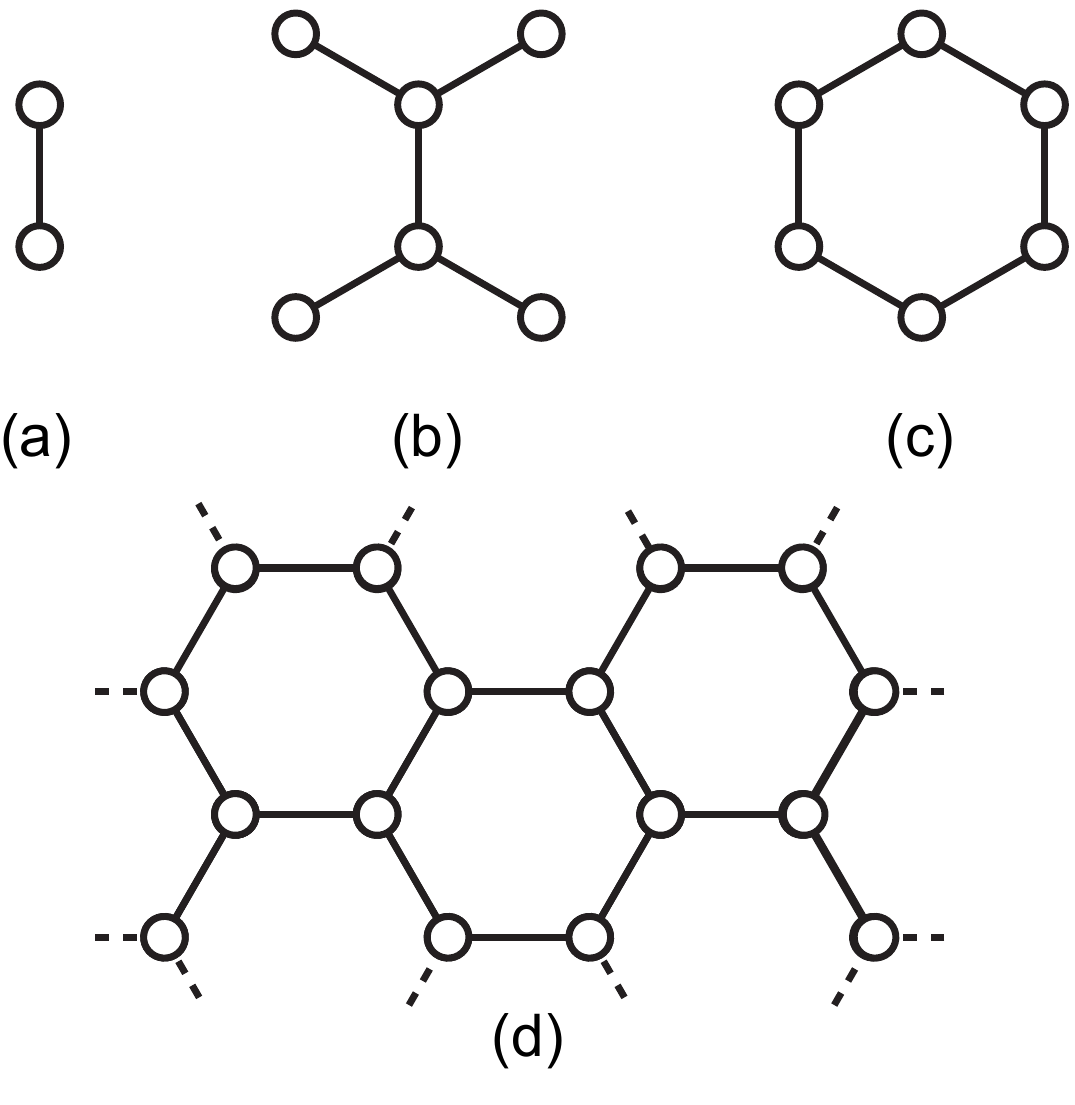}
\caption{\label{fig-clusters} Clusters employed in exact diagonalization studies for the extraction of magnetic parameters: (a) two-site cluster, (b) ``bridge'' cluster, and (c) ``hexagon'' cluster. The 16-site cluster (d) has been used for ED studies of the resulting magnetic models.}
\end{figure}

We show, in Table \ref{table_NaIr1}, a detailed comparison of nearest neighbour interactions computed at various levels of theory. The results of the ``exact'' second order perturbation theory (Ex. 2OPT) for $\Delta_n=0$, described in Sec. \ref{sec_2OPT}, are supplemented by parameters from exact diagonalization (ED) of the full Hamiltonian Eq. (\ref{eqn-1}) without (``no'') and with (``full'') inclusion of crystal field splitting (Appendix \ref{app_ED}). We have not included effects of $t_{2g}-e_g$ mixing explicitly; such effects have been estimated to shift the computed values $\sim 2$ meV, and therefore can be considered a small correction.\cite{Yamaji:2014be,Foyevtsova:2013uo} Calculations were performed on both 2-site and 6-site ``bridge'' clusters shown in Fig. \ref{fig-clusters}(a,b). The bridge cluster allows for estimation of higher order multi-site corrections to the nearest neighbour terms, and therefore represents the most accurate treatment available. In each case, it is important to choose clusters that explicitly retain any symmetry relevant to the interactions of interest. At present, larger cluster sizes are not practical due to the computational expense of exactly solving the full multi-orbital Hubbard problem. At all levels of theory, we find a dominant nearest neighbour Kitaev interaction $K_1 \sim -20$ meV, with both Heisenberg coupling $J_1$ and off-diagonal anisotropic exchange $\Gamma_1,\Gamma_1^\prime$ appearing roughly an order of magnitude smaller. In this way, qualitatively accurate results are already obtained at the level of ``exact'' second order perturbation theory with $\Delta_n=0$ for nearest neighbours. 

In contrast with the N2OPT calculation of Ref. \onlinecite{Yamaji:2014be}, the anisotropy between the X$_1$,Y$_1$, and Z$_1$-bonds is found to be relatively weak, and we do not find a large $\Gamma_1^{\prime xy}$ or $\zeta_1$ suggested to explain the observed zig-zag order. Instead, all off-diagonal interactions are $\sim 2$ meV in all methods employed. The effects of crystal field splitting (CFS) $\Delta_n$ is to shift all interactions by $\lesssim 2$ meV, as can be seen by comparing the 2-site ED results in the presence or absence of this term. Consistent with Ref. \onlinecite{Yamaji:2014be}, the CFS tends to slightly enhance $K_1$, while suppressing $J_1$, although this effect is relatively mild. Of somewhat greater importance is the renormalization of the nearest neighbour terms by multi-site corrections, which can be seen from comparisons of the 2-site and 6-site results. Indeed, as discussed above, the $j_{\text{eff}} = 1/2$ holes are rather delocalized, and higher order corrections captured on the bridge clusters can be quantitatively relevant.

Second and third neighbour interactions were also estimated from ED calculations on 6-site ``hexagon'' clusters shown in Fig. \ref{fig-clusters}(c). Full numerical results are given in Table \ref{table_NaIr2}. Contrary to the suggestion of Ref. \onlinecite{sizyuk2014importance}, we find no evidence for large second neighbour interactions, with bond-averaged values corresponding to $J_2 \sim +0.2$ meV, and $K_2 \sim -1.4$ meV. These interactions are generally suppressed due to the interference of the various second and third order hopping processes in Fig. \ref{fig-highorder}, which were not considered in Ref. \onlinecite{sizyuk2014importance}. In contrast, the calculated third neighbour interactions are large, and dominated by Heisenberg coupling $J_3 \sim +6.8$ meV, which greatly exceeds the estimate from N2OPT of +1.3 meV in Ref. \onlinecite{Yamaji:2014be}. The enhancement of this term in ED calculations results from the inclusion of all higher order contributions that are neglected at second order. Finally, we find no four-spin or six-spin ring-exchange terms exceeding 0.1 meV.

\begin{table}[t]
\caption{ \label{table_NaIr2} Complete magnetic interactions in meV for Na$_2$IrO$_3$ obtained by exact diagonalization on six-site bridge and hexagon clusters employing $U$ = 1.7 eV, $J_{\rm H}$ = 0.3 eV, $\lambda$ = 0.4 eV, and full crystal field terms $\Delta_n$. The largest terms are bolded. Site labels for $\mathbf{D}_{ij}$ refer to Fig. \ref{fig-labels}(a).}
\begin{ruledtabular}
\begin{tabular}{ccccccccc}
&Bond&$J_n$&$K_n$ &$\xi_n$& $\Gamma_n$&$\Gamma_n^\prime$&$\zeta_n$ \\
\hline \\[-2.0ex]
&X$_1$, Y$_1$ & -0.1 &{\bf -16.2} & +0.1 &+2.1 & -2.3&-0.1 \\
&Z$_1$ & +1.6 & {\bf -17.9} &$-$& -0.1 & -1.8 &$-$\\
\hline \\[-2.0ex]
&X$_2$, Y$_2$ & +0.2 & -1.6 & -0.1&+0.9 & 0.0 & 0.0  \\
&Z$_2$ & +0.1 & -1.2 &$-$& +0.6 & -0.3 &$-$\\
\hline \\[-2.0ex]
&X$_3$,Y$_3$&{\bf +6.7} & 0.0& 0.0 & -0.1 & 0.0  & -0.1\\
&Z$_3$&{\bf  +6.8} & +0.3 &$-$& -0.2 & -0.1&$-$\\
\end{tabular}
\end{ruledtabular}
\vspace{3mm}
\begin{ruledtabular}
\begin{tabular}{ccccc}
&Bond&Sites $(i,j)$&$\mathbf{D}_{ij}$ \\
\hline \\[-2.0ex]
&X$_2$ & (1, 3) , (4, 6) & (-0.1, -0.5, -0.5) \\
&Y$_2$ & (5, 1) , (2, 4) & (-0.5, -0.1, -0.5) \\
&Z$_2$ & (6, 2) , (3, 5) & (-0.2, -0.2, -0.1) \\
\end{tabular}
\end{ruledtabular}
\end{table}

\subsubsection{Minimal Model and Comparison to Experiment}

\begin{figure}[t]
\includegraphics[width=0.95\linewidth]{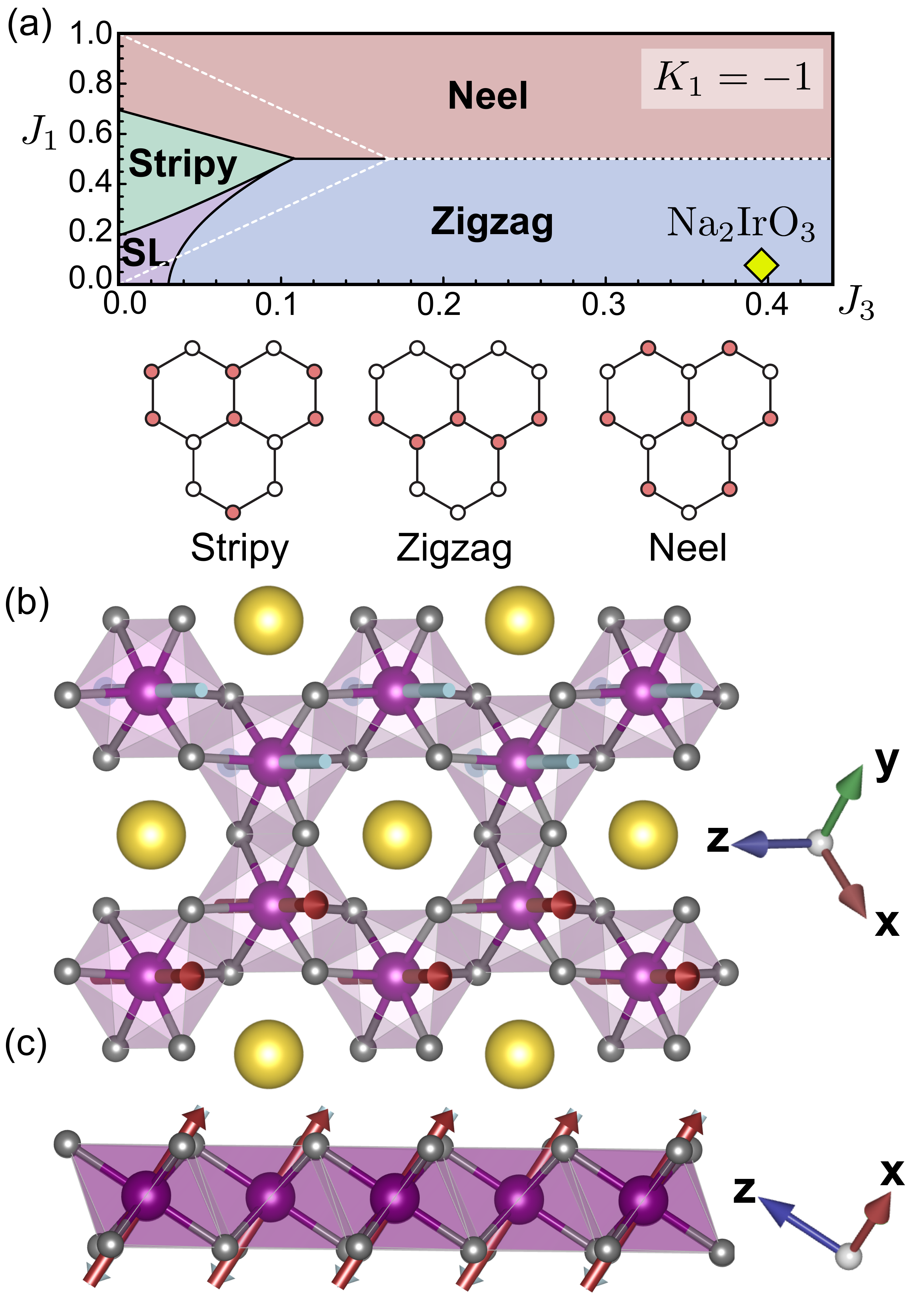}
\caption{\label{fig-Namag} (a) Phase diagram for the minimal model of Eq. (\ref{eqn-minNa}) obtained by ED on 16-site cluster in the parameter region relevant to the Na$_2$IrO$_3$; SL = Kitaev Spin Liquid. The white dashed lines indicate classical phase boundaries. (b,c) Predicted zigzag ground state and orientation of the ordered moments for Na$_2$IrO$_3$, viewed (b) along the cubic [111] direction, and (c) the cubic [$\bar{1}$10] direction. The moments are found to be $\perp b$-axis, nearly directly along the $\hat{x}+\hat{y}$ direction. }
\end{figure}
On the basis of the above calculations, we therefore suggest that the minimal model for zigzag order in Na$_2$IrO$_3$ is a $K_1$-$J_3$ model:
\begin{align}
\mathcal{H} = \sum_{\text{1st nn}}\left(J_1 \mathbf{S}_i \cdot \mathbf{S}_j + K_1 S_i^\gamma S_j^\gamma \right)+ \sum_{\text{3rd nn}} J_3\mathbf{S}_i\cdot\mathbf{S}_j \label{eqn-minNa}
\end{align}
with $J_1 \sim 0$, $J_3>0$, $K_1<0$, and $|J_3/K_1| \sim 0.4$. We show, in Fig. \ref{fig-Namag}, the phase diagram of this model obtained by exact diagonalization on the 16 site cluster shown in Fig. \ref{fig-clusters}(d); phase boundaries were identified from extrema of $\partial^2 E/(\partial J_n)^2$, where $E$ is the ground state energy. As can be seen, robust collinear zigzag order emerges naturally at large $J_3/J_1$, as a means of satisfying all third neighbour interactions.\cite{kimchi2011kitaev, Katukuri:2014iq} In the limit $|K_1|\gg |J_1|$, the zigzag ground state can be expected even for $J_3$ much smaller than that computed in this work, suggesting this result should be robust even for different choices of model parameters $J_H,U,\lambda$. In the real material, the further details of the bond anisotropies and off-diagonal anisotropic interactions determine the specific ordering wave vector $\mathbf{q}$ and ordered moment direction $\mathbf{n}$. For the computed interactions, the classical energy is minimized for $\mathbf{q}||$Z$_1$-bond (the crystallographic $b$-axis), while the $\Gamma_1^{xy}$ term for the X$_1$ and Y$_1$ bonds ensures that $\mathbf{n}$ is oriented at 45$^\circ$ between the cubic $x$ and $y$ axes, as observed in experiment.\cite{HwanChun:2015ed} The small $\Gamma_1^\prime < 0$ may also stabilize the zigzag state.\cite{Rau:2014wc} The absence of significant bond-anisotropy (i.e. $K_1^z \approx K_1^{xy}$) is consistent with the observation of near $C_3$ symmetry of the observed magnetic fluctuations above $T_N$.\cite{HwanChun:2015ed} We may further estimate the Weiss constant given a particular orientation of the magnetic field $\hat{h}(\theta,\phi)$ via:
\begin{align}
\Theta(\theta,\phi) =& \ -\frac{S(S+1)}{3k_b}\sum_j \hat{h}\cdot \mathbf{J}_{ij} \cdot \hat{h}
\end{align}
where the summation is over all bonds connected to a given site $i$. We approximate the isotropic Weiss constant as the average value:
\begin{align}
\Theta_{iso} \sim& \ \int \Theta(\theta,\phi) \sin \theta \  d\theta \ d\phi \\
\sim & \ -\frac{1}{4k_b}\left(3J_1+6J_2+3J_3+K_1+2K_2\right)
\end{align}
which is independent of all off-diagonal $\Gamma_1,\Gamma_1^\prime$ terms. For the interactions computed in this work, we find $\Theta_{iso} < 0$, and $\Theta_b<\Theta_a<\Theta_{c^*}$, (i.e. $\chi_{ab}<\chi_{c^*}$), consistent with experiment.\cite{Singh:2010gg} Indeed, if it is assumed that Na$_2$IrO$_3$ is relatively close to the ferromagnetic Kitaev limit at the nearest neighbour level, then the antiferromagnetic $\Theta_{iso} \sim -116$ K can only be explained by additional long-range antiferromagnetic coupling such as $J_3$. Similar conclusions were reached via analysis of $\chi$ in Ref. \onlinecite{kimchi2011kitaev}. We leave for future work a full comparison between the computed interactions and the inelastic neutron scattering results of Ref. \onlinecite{choi2012spin}.
Given that the interactions in Na$_2$IrO$_3$ are strongly frustrated at the nearest neighbour level by the large Kitaev terms, it is not clear that linear spin-wave theory provides an accurate description of the excitation spectrum, as discussed in Ref.~\onlinecite{Yamaji:2016ws}.

\subsection{$\alpha$-RuCl$_3$}
\subsubsection{Introduction}

As with Na$_2$IrO$_3$, the honeycomb trihalide $\alpha$-RuCl$_3$ displays zigzag order below $T_N \sim 7-14$ K.\cite{Johnson:2015vb,nagler2016new} However, observation of a ferromagnetic Weiss constant $\Theta_{iso} \sim +40$ K, and a reversed susceptibility anisotropy, (i.e. $\chi_{ab} > \chi_{c}$) suggest a different character to the magnetic interactions.\cite{kobayashi1992moessbauer,Sears:2015ku} Very early structural studies indicated a highly symmetric $P3_112$ space group,\cite{Fletcher:1967to} with nearly isotropic edge-sharing RuCl$_6$ octahedra, implying relatively weak crystal field splitting. It was thus argued that Kitaev physics may be realized in the lighter $4d^5$ ruthenium, despite weaker SOC ($\lambda_{Ru} \sim 0.15$ eV).\cite{Kim:2015iq} The case for large and frustrated anisotropic interactions was strengthened by analysis of neutron scattering data, which was fit by $K_1 \sim +7$ meV, and $J_1 \sim -4$ meV.\cite{Banerjee:2015wi} However, early concerns\cite{brodersen1968struktur} over the correct identification of the space group were recently raised again;\cite{Kubota:2015gu, Kim:2015ue,nagler2016new} in the $\alpha$-RuCl$_3$ structure, the weak interactions between hexagonal layers result in significant structural defects, which complicates structural solution. Recent detailed studies found instead monoclinic $C2/m$ packing analogous to $A$$_2$IrO$_3$ for a single crystals shown to exhibit single transitions to zigzag order.\cite{Johnson:2015vb,nagler2016new} These results are in contrast with previous samples exhibiting multiple magnetic transitions,\cite{Kubota:2015gu} which may be induced by physical distortion of the samples.\cite{nagler2016new} Not surprisingly, the revised $C2/m$ crystal structures have significantly enhanced distortion of the RuCl$_6$ octahedra when compared with the assumed $P3_112$ structure, prompting a reanalysis of the magnetic interactions.

The first study on the magnetic interactions in $\alpha$-RuCl$_3$ was performed on the $P3_112$ structure, and employed the projective expressions of Ref. \onlinecite{rau2014generic}, together with hopping integrals from DFT.\cite{Kim:2015iq} The authors suggested $J_1 \sim -12$, $K_1 \sim +17$, and $\Gamma_1\sim+12$ meV, correctly placing the material in a region expected to display zigzag order, and emphasized the importance of $t_{2g}-e_g$ mixing, which enhances the $K>0$ and $J<0$.\cite{khaliullin2005orbital} It is worth noting, however, that the latter conclusion was reached after neglecting Coulomb repulsion between the $t_{2g}$ and $e_{g}$ orbitals, and therefore deserves reevaluation. Subsequent analysis was also performed on theoretical $C2/m$ structures for $\alpha$-RuCl$_3$ obtained by relaxation within DFT.\cite{Kim:2015ue} This analysis found instead ferromagnetic Kitaev coupling $K_1<0$, placing emphasis on the structural dependence of such interactions. In this section, we provide a detailed reanalysis of the magnetic interactions for the original $P3_112$ and new experimental $C2/m$ structure of Ref. \onlinecite{Johnson:2015vb} in order to address the possible variations to the in-plane interactions that might occur due to structural distortions.

\subsubsection{Calculations and Discussion}

\begin{table}[t]
\caption{ \label{table_RuCl1} Complete magnetic interactions in meV for the $C2/m$ structure of $\alpha$-RuCl$_3$ from Ref. \onlinecite{Johnson:2015vb} obtained by exact diagonalization on six-site bridge and hexagon clusters employing $U$ = 3.0, $J_{\rm H}$ = 0.6, $\lambda$ = 0.15 eV, and full crystal field terms $\Delta_n$. The largest terms are bolded. Site labels for $\mathbf{D}_{ij}$ refer to Fig. \ref{fig-labels}(a).}
\begin{ruledtabular}
\begin{tabular}{ccccccccc}
&Bond&$J_n$&$K_n$ &$\xi_n$& $\Gamma_n$&$\Gamma_n^\prime$&$\zeta_n$ \\
\hline \\[-2.0ex]
&X$_1$, Y$_1$ & -1.4 &{\bf -7.5} & +0.2 &{\bf +5.9 }& -0.8&+0.2 \\
&Z$_1$ & -2.2 & {\bf -5.0} &$-$&\bf{ +8.0} & -1.0 &$-$\\
\hline \\[-2.0ex]
&X$_2$, Y$_2$ & -0.1 & -0.6 & +0.1&+0.6 & +0.6 & +0.1  \\
&Z$_2$ & +0.1 & -0.9 &$-$& +0.6 & +0.3 &$-$\\
\hline \\[-2.0ex]
&X$_3$,Y$_3$&{\bf +3.0} &-0.1& 0.0 & -0.1 &-0.1  & -0.1\\
&Z$_3$&{\bf  +2.4} & +0.3 &$-$& -0.1 & -0.1&$-$\\
\end{tabular}
\end{ruledtabular}
\vspace{3mm}
\begin{ruledtabular}
\begin{tabular}{ccccc}
&Bond&Sites $(i,j)$&$\mathbf{D}_{ij}$ \\
\hline \\[-2.0ex]
&X$_2$ & (1, 3) , (4, 6) & (-0.3, -0.5, -0.5) \\
&Y$_2$ & (5, 1) , (2, 4) & (-0.5, -0.3, -0.5) \\
&Z$_2$ & (6, 2) , (3, 5) & (-0.4, -0.4, -0.1) \\
\end{tabular}
\end{ruledtabular}
\end{table}

We show in Table \ref{table_RuCl1} the nearest neighbour interactions extracted from calculations on 6-site bridge clusters for the $C2/m$ structure. In order to avoid discussion of the local symmetry-allowed interactions for the $P3_112$ structure, we present only the bond-averaged values computed on the six-site bridge clusters: ($J_1,K_1,\Gamma_1,\Gamma_1^\prime$) = (-5.5, +7.6, +8.4, +0.2) meV, respectively. In contrast, the $C2/m$ structure displays a somewhat smaller Heisenberg coupling, and a ferromagnetic Kitaev term: ($J_1,K_1,\Gamma_1,\Gamma_1^\prime$) = (-1.7, -6.7, +6.6, -0.9) meV, respectively. For both structures, we find a ferromagnetic Heisenberg coupling $J_1 < 0$, and a dominant $\Gamma_1>0$ interaction, which results from the large metal-metal hopping $t_1,t_3$, consistent with the previous studies.\cite{Kim:2015ue,Kim:2015iq} We also note a somewhat significant bond-anisotropy for the $C2/m$ structure of Ref. \onlinecite{Johnson:2015vb}, with $K_1^{xy}<K_1^z$ and $\Gamma_1^{xy} <\Gamma_1^z$, which results primarily from anisotropy in the $t_3,t_3^\prime$ hopping integrals. As with Na$_2$IrO$_3$, we find no large second neighbour interactions, with all terms $< 1$ meV. However, both the $P3_112$ and $C2/m$ structures of $\alpha$-RuCl$_3$ display significant third neighbour Heisenberg coupling arising from high-order processes: $J_3 \sim 2.3$ meV for $P3_112$ and $J_3 \sim 2.7$ meV for $C2/m$ structures. These values are a full order of magnitude larger than previous 2OPT estimates.\cite{Kim:2015iq} We note that consideration of the $C2/m$ structure of Ref. \onlinecite{nagler2016new} suggests a somewhat reduced bond-anisotropy, but no significant modification of the computed interactions of Table \ref{table_RuCl1}. 

\subsubsection{Minimal Model and Comparison to Experiment}

\begin{figure}[t]
\includegraphics[width=0.95\linewidth]{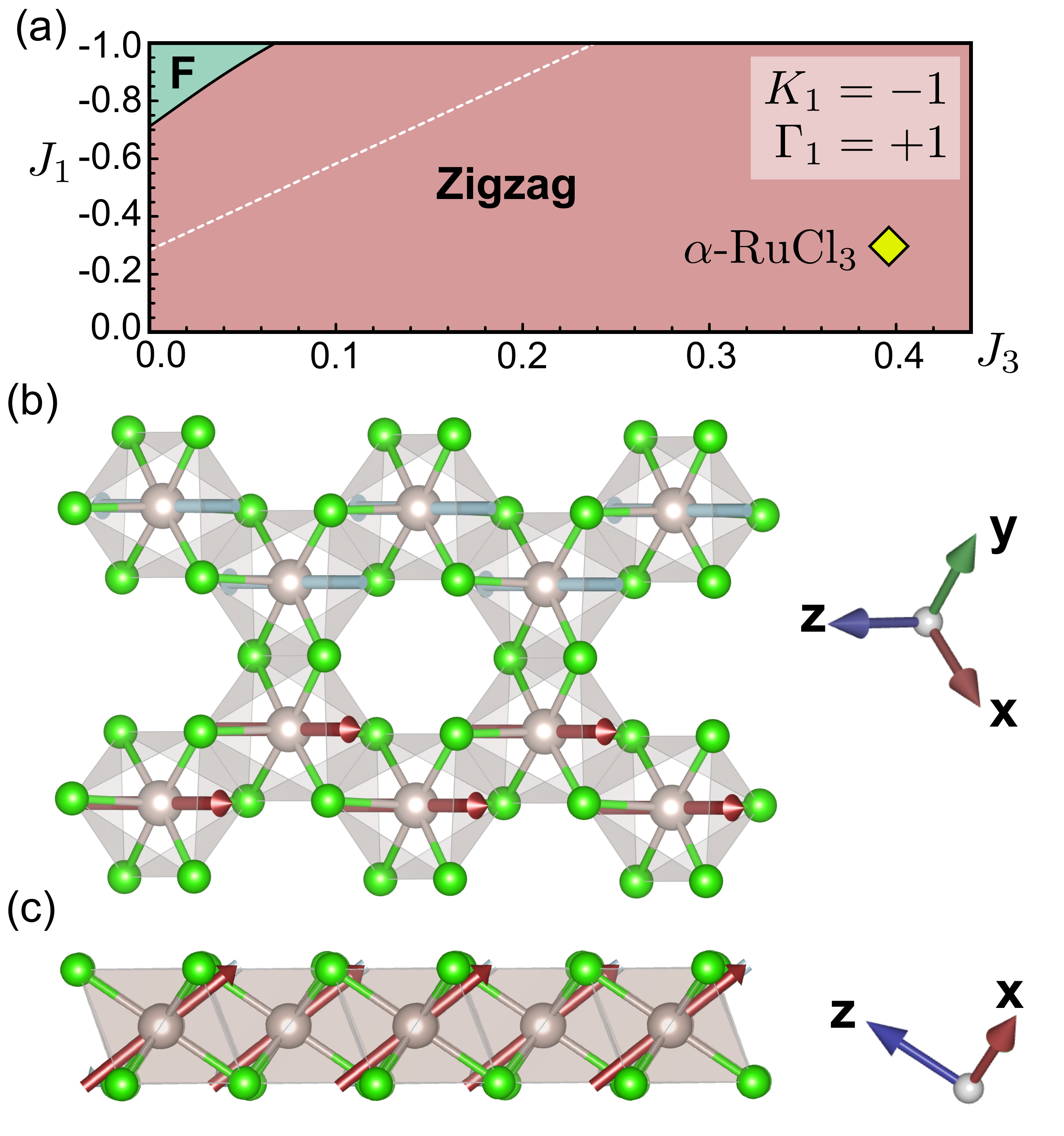}
\caption{\label{fig-Rumag} (a) Phase diagram of the minimal model of Eq. (\ref{eqn-minRu}) obtained by ED on 16-site cluster in the parameter region relevant to the $C2/m$ structure of $\alpha$-RuCl$_3$; F = bulk ferromagnetic order. The white dashed line indicates the classical phase boundary. (b,c) Predicted zigzag ground state and orientation of the ordered moments for the $C2/m$ structure of $\alpha$-RuCl$_3$, viewed (b) along the cubic [111] direction, and (c) the cubic [$\bar{1}$10] direction. The moments are found to be $\perp b$-axis, nearly directly along the $\hat{x}+\hat{y}$ direction, but tilted 106$^\circ$ from the cubic $z$-axis. }
\end{figure}
On the basis of the computed interactions, we identify the relevant terms as: $(J_1,K_1,\Gamma_1,J_3)$, with bond averaged values $(-1.7,-6.7,+6.6,+2.7)$ meV. This finding implies a minimal model with $\Gamma_1=-K_1$:
\begin{align}
\nonumber \mathcal{H}= \sum_{\text{1st nn}}& \ J_1 \ \mathbf{S}_i \cdot \mathbf{S}_j + K_1(S_i^\gamma S_j^\gamma -S_i^\alpha S_j^\beta - S_i^\beta S_j^\alpha ) \\ & \ +\sum_{\text{3rd nn}}J_3 \ \mathbf{S}_i \cdot \mathbf{S}_j \label{eqn-minRu}
\end{align}
where $\{\alpha,\beta,\gamma\}=\{x,y,z\}$ for the Z$_1$ bond, for example. This Hamiltonian can be brought into a more convenient form via a 45$^\circ$ rotation around each local $\gamma$ axis to yield:
\begin{align}
\mathcal{H}= \sum_{\text{1st nn}}& \ A_1 \mathbf{S}_i \cdot \mathbf{S}_j +B_1 S_i^\delta S_j^\delta +\sum_{\text{3rd nn}}J_3 \mathbf{S}_i \cdot \mathbf{S}_j \label{eqn-33}
\end{align}
where $A_1=J_1+K_1\sim -8.4$ meV, $B_1=-2K_1\sim +13.4$ meV, with $A_1/B_1 \sim -0.63$, and the $\hat{\delta}$-axis is a bond-dependent direction given by:
\begin{align}
\hat{\delta} = \left\{\begin{array}{c}\frac{1}{\sqrt{2}}(0,1,1) \ \ \ \ \  \text{X$_1$ bond} \\ \frac{1}{\sqrt{2}}(1,0,1) \ \ \ \ \  \text{Y$_1$ bond}  \\ \frac{1}{\sqrt{2}}(1,1,0) \ \ \ \ \  \text{Z$_1$ bond} \end{array} \right.
\end{align}
That is, for each bond, $\hat{\delta}$ is parallel to a vector joining the two bridging Cl atoms. It is worth noting that, while the anisotropic terms in Eq. (\ref{eqn-33}) take a Kitaev-like form, the $\hat{\delta}$-axes are not orthogonal to one another, but rather intersect at an angle of 60$^\circ$, significantly reducing frustration. Nonetheless, at the classical level, this model permits the same ground states as the conventional Heisenberg-Kitaev model with $(J_1<0, K_1>0,J_3>0)$: bulk ferromagnetism is found for $(A_1+3J_3)/B_1 \lesssim -0.64$, while zigzag order appears for $(A_1+3J_3)/B_1 \gtrsim -0.64$. We show, in Fig. \ref{fig-Rumag}(a), the phase diagram of this model obtained by exact diagonalization on the 16 site cluster shown in Fig. \ref{fig-clusters}(d). As with Na$_2$IrO$_3$, the observed zigzag order in $\alpha$-RuCl$_3$ is uniquely selected by the large $J_3>0$ coupling. Given this, the zigzag state is expected to be stable against even large structural distortions, but variations in $T_N$ likely arise from a modulation of both the interplane and intraplane interactions. The effects of the mild bond-anisotropy on the ordered state are unclear, but deserve further study. Employing all computed interactions for the $C2/m$ structure of Ref. \onlinecite{Johnson:2015vb}, we find for the collinear zigzag state that the classical energy is minimized for an ordering wave vector $\mathbf{q}||$Z$_1$ bond, although the states with $\mathbf{q}||$X$_1$,Y$_1$ bonds are not significantly different in energy. For $\mathbf{q}||$Z$_1$, the moments are oriented close to the cubic $(110)$ direction, but inclined 106$^\circ$ degrees from the cubic $z$-axis. The moments are thus predicted to make an angle of $\sim 30^\circ$ with the $ab$-plane, which is consistent with the predictions of Ref. \onlinecite{PhysRevB.92.024413} for significant $|\Gamma_1/(K_1+\Gamma_1^\prime)|$. For the computed interactions, we correctly find a ferromagnetic $\Theta_{iso}\sim -(3A_1+B_1+3J_3)/4k_b>0$, which results from the $J_1,K_1<0$. We further find $\Theta_{c^*} < 0$ and $\Theta_a,\Theta_b > 0$, in agreement with experiment.\cite{kobayashi1992moessbauer,Sears:2015ku}

\subsection{$\alpha$-Li$_2$IrO$_3$ \label{sec_Li}}
\subsubsection{Introduction}

The final material addressed in this work is the $\alpha$-phase of Li$_2$IrO$_3$, which is considered to be isostructural to the Na analogue, and similarly exhibits magnetic order below $T_N = 15$ K.\cite{singh2012relevance} However, evidence for a different character of this order can be seen in the strong suppression of $T_N$ upon Na$\rightarrow$Li substitution of Na$_2$IrO$_3$.\cite{Manni2014,Cao:2013bz} Indeed, initial powder neutron experiments on $\alpha$-Li$_2$IrO$_3$ suggested an incommensurate phase with a $\mathbf{q}$-vector in the first Brillouin zone,\footnote{R. Coldea, talk at the SPORE13 conference held at MPI-
PKS, Dresden, 2013; S. Choi, talk at the APS March meet- ing, Denver, CO, 2014; R. Coldea, talk at the DPG-focus session, spring meeting of the German Physical Society, Dresden, 2014.} prompting several theoretical proposals for the origin of such order on the basis of simplified models. Very recently, the  availability of single crystals allowed for detailed X-ray and neutron studies, which reveal a complex counter-rotating spiral phase with $\mathbf{q}||a^*$, and $|\mathbf{q}| = q \sim 0.32$,\cite{radu2016new} similar to states observed in the 3D hyperhonecomb $\beta$- and stripyhoneycomb $\gamma-$ phases of Li$_2$IrO$_3$.\cite{biffin2014noncoplanar,biffin2014unconventional}
Regarding $\alpha$-Li$_2$IrO$_3$, intuition gained from studies of the pure Heisenberg $(J_1,J_2)$ model led the authors of Ref. \onlinecite{Reuther:2014ku} to suggest the importance of second neighbour interactions. Indeed, spiral phases are found in the $(J_1,K_1,J_2,K_2)$ model,\cite{Reuther:2012bb,Reuther:2014ku,sizyuk2014importance} which are presumably connected to similar states in the pure Heisenberg $(J_1,J_2)$ case. Evidence for significant long-range interactions for $\alpha$-Li$_2$IrO$_3$ had been argued for based on analysis of the magnetic susceptibility,\cite{kimchi2011kitaev} and nonmagnetic doping studies.\cite{manni2014effect} Alternatively, incommensurate order has also been shown to emerge in pure nearest neighbour $(J_1,K_1,\Gamma_1,\Gamma_1^\prime)$ models provided $\Gamma_1,\Gamma_1^\prime$ are large.\cite{Rau:2014wc,rau2014generic} Along this line, the authors of Ref. \onlinecite{PhysRevB.92.024413} suggested that an appropriate starting point may be $K_1 \sim -\Gamma_1 \sim -10$ meV, and $J_1 \sim \Gamma_1^\prime \sim 0$. Finally, a third scenario was formulated in Ref. \onlinecite{Kimchi:2015ky}, where the effects of bond-anisotropy were considered for a nearest neighbour model. The authors introduced a dipolar coupling $I_c<0$ for the Z$_1$ bonds only, equivalent to the choice: $J_1^{xy} = J$, and $K_1^{xy} = K$ for X$_1$ and Y$_1$ bonds, and: $J_1^z = J + I_c/2, \ K_1^z = K - I_c/2, \ \Gamma_1^z = -I_c/2$
for Z$_1$ bonds. They further suggested an example parameter set consistent with experiment: $(J_1^z,K_1^z,\Gamma_1^z,J_1^{xy},K_1^{xy})$ $\sim (-2,-10,+2,+1,-12)$ meV. Given this significant range of proposals, insight from {\it ab-initio} methods is vital. However, a significant challenge for {\it ab-initio} modelling of $\alpha$-Li$_2$IrO$_3$ arises from uncertainty in the crystal structure, which has only been reported from powder X-ray analysis due to the unavailability, until recently, of highly ordered single crystal samples.\cite{Gretarsson2013} As pointed out in section \ref{sec_hop}, this structure exhibits significant anisotropy in the hopping integrals comparing different bonds. The only reported {\it ab-initio} study to date extracted nearest neighbour magnetic interactions from MRCI state energies on 2-site clusters starting from this experimental structure.\cite{Nishimoto:2014vw} The results reflected the significant bond-anisotropy, with Z$_1$ bonds displaying large ferromagnetic Heisenberg coupling: $(J_1^z,K_1^z) = (-19,-6)$ meV, while the X$_1$ and Y$_1$ bonds were dominated by a ferromagnetic Kitaev exchange: $(J_1^{xy},K_1^{xy})=(+1,-12)$ meV. Given these interactions, the authors suggested that spiral phases could emerge if supplemented by a large $J_2$. However, such states are inconsistent with the observed order.\cite{radu2016new} More detailed studies are currently lacking. Given that the degree of bond-anisotropy in the real material may be less than the published powder X-ray structure, we have also considered an alternative theoretical structure obtained by relaxation of the experimental atomic positions within DFT.\cite{Manni2014}

\subsubsection{Calculations and Discussion}

\begin{table}[t]
\caption{ \label{table_LiIr1} Complete magnetic interactions in meV for $\alpha$-Li$_2$IrO$_3$  obtained by exact diagonalization on six-site bridge and hexagon clusters employing $U$ = 1.7 eV, $J_{\rm H}$ = 0.3 eV, $\lambda$ = 0.4 eV, and full crystal field terms $\Delta_n$. The largest terms are bolded. Site labels for $\mathbf{D}_{ij}$ refer to Fig. \ref{fig-labels}(a).}
\begin{ruledtabular}
\begin{tabular}{ccccccccc}
\multicolumn{8}{c}{Experimental Structure}\\
\hline \\[-2.0ex]
&Bond&$J_n$&$K_n$ &$\xi_n$& $\Gamma_n$&$\Gamma_n^\prime$&$\zeta_n$ \\
\hline \\[-2.0ex]
&X$_1$, Y$_1$ & {\bf -1.0}&\bf -13.0 & -0.1 &\bf +6.6& -0.4&+0.6 \\
&Z$_1$ & \bf -4.6 & \bf -4.2 &$-$&\bf +11.6 &\bf  -4.3 &$-$\\
\hline \\[-2.0ex]
&X$_2$, Y$_2$ & +0.9 &\bf  -2.9 & +1.3&\bf  +3.0 & +1.3 & +0.4  \\
&Z$_2$ & -0.9 &+0.1 &$-$& +1.5 & -1.6 &$-$\\
\hline \\[-2.0ex]
&X$_3$,Y$_3$&{\bf +4.7} &-0.2& -0.1 & 0.0 &0.0  & -0.1\\
&Z$_3$&{\bf +4.4}& +0.4 &$-$& -0.1 & -0.1&$-$\\
\end{tabular}
\end{ruledtabular}
\vspace{3mm}
\begin{ruledtabular}
\begin{tabular}{ccccc}
&Bond&Sites $(i,j)$&$\mathbf{D}_{ij}$ \\
\hline \\[-2.0ex]
&X$_2$ & (1, 3) , (4, 6) &\bf  (-1.5, -3.2, -2.3) \\
&Y$_2$ & (5, 1) , (2, 4) &\bf  (-3.2, -1.5, -2.3) \\
&Z$_2$ & (6, 2) , (3, 5) & (-0.2, -0.2, 0.0) \\
\end{tabular}
\end{ruledtabular}
\vspace{3mm}
\begin{ruledtabular}
\begin{tabular}{ccccccccc}
\multicolumn{8}{c}{Relaxed Structure}\\
\hline \\[-2.0ex]
&Bond&$J_n$&$K_n$ &$\xi_n$& $\Gamma_n$&$\Gamma_n^\prime$&$\zeta_n$ \\
\hline \\[-2.0ex]
&X$_1$, Y$_1$ & \bf -2.5 &\bf -9.8 & 0.0 &\bf +8.7& -0.8&+0.1 \\
&Z$_1$ & \bf -3.1 &\bf  -6.3 &$-$&\bf +9.4 & -0.1&$-$\\
\hline \\[-2.0ex]
&X$_2$, Y$_2$ & +0.5 & \bf -3.8 & +1.0&\bf +3.4 & +0.5 & +0.1  \\
&Z$_2$ &+0.2 &\bf  -3.6 &$-$& \bf +3.3 & -0.6 &$-$\\
\hline \\[-2.0ex]
&X$_3$,Y$_3$&\bf +6.0&-0.1& -0.1 & 0.0 &-0.1  & -0.1\\
&Z$_3$&\bf +5.9&+0.2 &$-$& -0.1 & -0.1&$-$\\
\end{tabular}
\end{ruledtabular}
\vspace{3mm}
\begin{ruledtabular}
\begin{tabular}{ccccc}
&Bond&Sites $(i,j)$&$\mathbf{D}_{ij}$ \\
\hline \\[-2.0ex]
&X$_2$ & (1, 3) , (4, 6) &\bf  (-0.3, -1.9, -1.4) \\
&Y$_2$ & (5, 1) , (2, 4)  & \bf (-1.9, -0.3, -1.4) \\
&Z$_2$ & (6, 2) , (3, 5) &\bf  (-1.2, -1.2, +0.1) \\
\end{tabular}
\end{ruledtabular}
\end{table}

Nearest neighbour interactions computed on 6-site bridge clusters are shown in Table \ref{table_LiIr1} for both the experimental\cite{Gretarsson2013} and relaxed\cite{Manni2014} structures. In both cases, we find ferromagnetic Heisenberg and Kitaev nearest neighbour interactions ($J_1^z,J_1^{xy},K_1^z,K_1^{xy}<0$), with significant bond-anisotropy appearing for the experimental structure, consistent with the results of Ref. \onlinecite{Nishimoto:2014vw}. The details of the bond-anisotropy (i.e. $K_1^z>K_1^{xy}$, $J_1^z<J_1^{xy}$, and $\Gamma_1^z>\Gamma_1^{xy}$) are consistent with the effects of the dipolar coupling model introduced in Ref. \onlinecite{Kimchi:2015ky}, although the latter model is considerably simplified compared to the computed interactions. As might be expected from the significant direct hopping $t_3$ (section \ref{sec_hop}), we find large off-diagonal $\Gamma_1^z, \Gamma_1^{xy}$ terms, similar to the case of $\alpha$-RuCl$_3$. The large computed $t_4 = -124.5$ meV for the experimental structure also leads to $\Gamma_1^{\prime z}=-4.3$ meV along the Z$_1$ bond only. Both the experimental and relaxed structures display similar bond-average values $J_1 \sim -3, K_1\sim -8,\Gamma_1\sim+9$ meV, while the latter structure has reduced bond-anisotropy, as might be expected. These average values are consistent with the predictions of Ref. \onlinecite{PhysRevB.92.024413}, and suggest that $\alpha$-Li$_2$IrO$_3$ is far away from the Kitaev limit at the nearest neighbour level, contrary to initial suggestions.\cite{singh2012relevance}

Second and third neighbour interactions for $\alpha$-Li$_2$IrO$_3$ were estimated from ED calculations on 6-site hexagon clusters, as shown in Table \ref{table_LiIr1}. Unlike Na$_2$IrO$_3$ and $\alpha$-RuCl$_3$, we find large second neighbour interactions. For the experimental structure, these interactions are particularly strong along the X$_2$ and Y$_2$ bonds, for which we estimate a large Dzyaloshinskii-Moriya term $|\mathbf{D}_2^{xy}| > 4$ meV that has not been previously considered in the literature. It can be shown that this term does {\it not} arise at $\mathcal{O}(t^2)$, as all second order contributions rely on intraorbital (i.e. $d_{xy}\rightarrow d_{xy}$) second neighbour hopping, which is very small in all honeycomb materials (see Appendices \ref{app_hop},\ref{app_terms}). Instead, the DM-term is due to third-order processes $\propto t_4t_2$, and can be expected in any material with significant $t_4$. We also note that the $\mathbf{\Gamma}_2^X$ tensor cannot be decomposed into pure second neighbour $(J_2,K_2)$ form that has been considered previously,\cite{Reuther:2014ku,sizyuk2014importance} but rather contains significant off-diagonal terms. This observation calls into question the applicability of previous theoretical studies of simplified models. For the relaxed structure, we also find large second neighbour terms, but both the off-diagonal $\Gamma_2^\prime$ and DM terms are reduced compared with the experimental structure. Finally, for both the experimental and relaxed structures, we find large third neighbour Heisenberg coupling of $J_3 =$ +4.6 and +5.9 meV, respectively.

\subsubsection{Minimal Model and Comparison to Experiment}

\begin{figure}[t]
\includegraphics[width=0.98\linewidth]{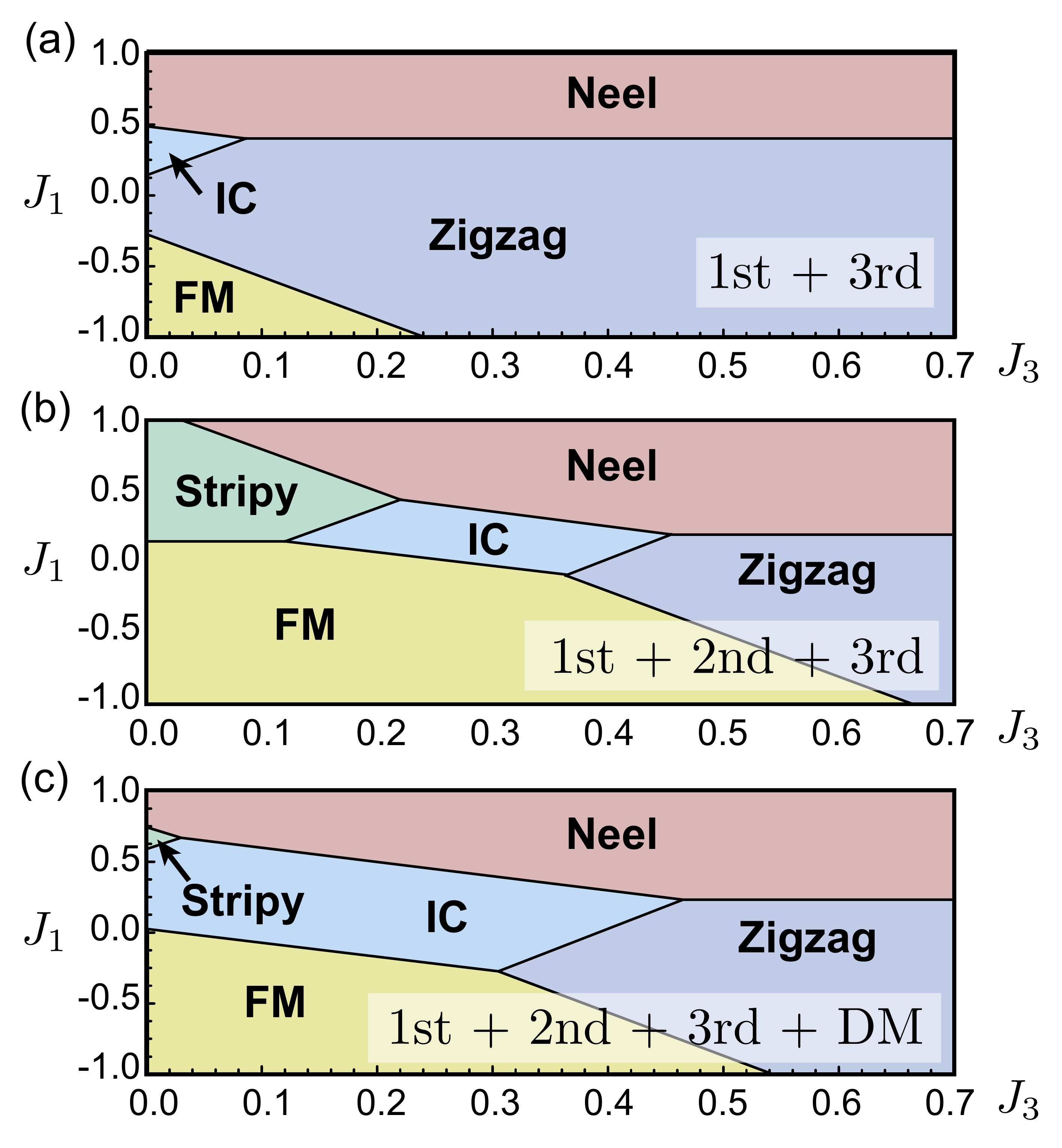}
\caption{\label{fig-liphase} Classical phase diagrams for model eq'n (\ref{eqn-35}) for $\alpha$-Li$_2$IrO$_3$ with increasingly realistic interactions. IC = incommensurate spiral order. (a) Including only first and third neighbour interactions; $A_2=B_2=D=0$, and $B_1=+2$, i.e. $K_1=-\Gamma_1=-1$. (b) With $D=0$, and with the ratio of other interactions set to reproduce the experimental $q$ and $A_a/A_{c^*}$; ($B_1=+2$, $B_2/B_1 = 0.3$, $A_2/B_2 = 0.5$). (c) With $D=0.15$, and with the ratio of other interactions set to reproduce the experimental $q$ and $A_a/A_{c^*}$; ($B_1=+2$, $B_2/B_1 = 0.2$, $A_2/B_2 = 0.5$).}
\end{figure}

\begin{figure}[t]
\includegraphics[width=0.98\linewidth]{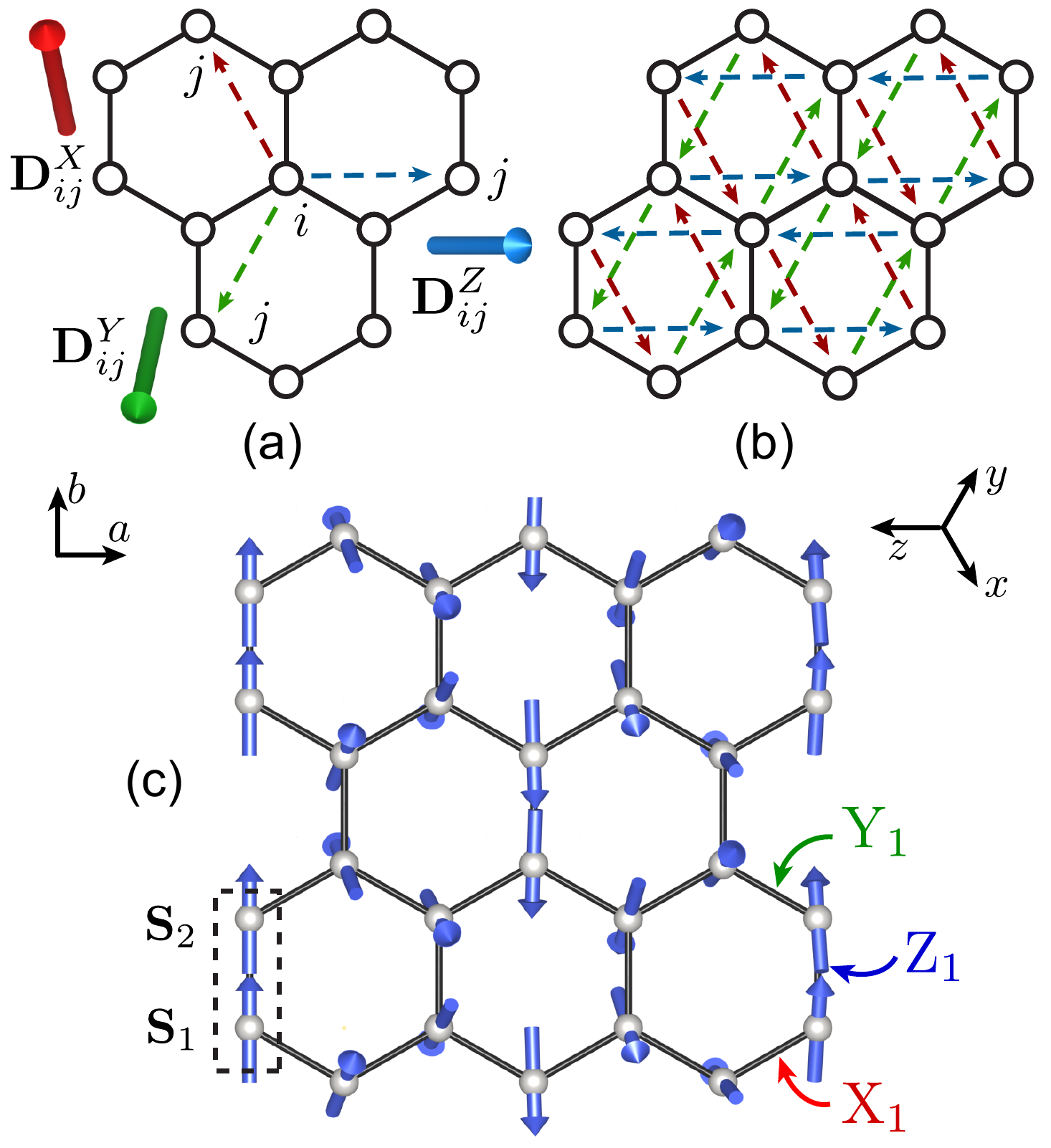}
\caption{\label{fig-linet} (a) Orientation of computed second neighbour DM vectors for the relaxed structure of $\alpha$-Li$_2$IrO$_3$, with sense of each interaction indicated by dashed arrows. (b) Network of interactions in the full lattice. Within a given sublattice, the DM vectors are uniform, and will therefore tend to promote incommensurate states. (c) Classical ground state of the minimal model of eq'n (\ref{eqn-35}) with $(A_1,B_1,A_2,B_2,J_3) \sim$ (-10.7, +24, -3.3,+7.0,+3.6) consistent with the experimental structure of Ref. \onlinecite{radu2016new}. }
\end{figure}

Given the structural uncertainty and indication of many relevant interactions, discussion of all details of the spiral order in $\alpha$-Li$_2$IrO$_3$ remains challenging. While previous works considering only nearest neighbour terms have argued for the importance of bond-anisotropy,\cite{Kimchi:2015ky} the results of the previous section imply that long-range second and third neighbour couplings may also play a significant role. Our approach for discussing the effects of such further neighbour terms is therefore to first consider phases that might emerge from a bond-averaged model in the vicinity of the computed parameters. On the basis of the above calculations, we therefore suggest that an appropriate starting point for $\alpha$-Li$_2$IrO$_3$ may be $(J_1,K_1,\Gamma_1,K_2,\Gamma_2,|\mathbf{D}_2|,J_3) = (-3,-8,+9,-4,+3,+3,+6)$ meV, strongly emphasizing the importance of long-range interactions. The finding that $K_1 \approx -\Gamma_1$ and $K_2 \approx -\Gamma_2$ suggests a minimal model similar to that of $\alpha$-RuCl$_3$:
\begin{align}
 \label{eqn-35}\mathcal{H}=& \ \sum_{\text{1st nn}} A_1 \mathbf{S}_i \cdot \mathbf{S}_j +B_1 S_i^\delta S_j^\delta +\sum_{\text{3rd nn}}J_3 \mathbf{S}_i \cdot \mathbf{S}_j \\ & \ +\sum_{\text{2nd nn}} \ A_2 \mathbf{S}_i \cdot \mathbf{S}_j +B_2 S_i^\delta S_j^\delta +\mathbf{D}_{ij} \cdot \mathbf{S}_i\times \mathbf{S}_j\nonumber
\end{align}
For simplicity, we consider a second neighbour DM-interaction of $C3$ symmetry: $\mathbf{D}_2^X=(0,-D,-D)$,  $\mathbf{D}_2^Y=(-D,0,-D)$, and $\mathbf{D}_2^Z=(-D,-D,0)$, which is inspired by the results for the relaxed structure. Our starting point for discussion is therefore $(A_1,B_1,A_2,B_2,D,J_3) \sim$ (-11, +16, -3, +7, +1.5, +6) meV. In this model, we expect that the incommensurate magnetic order is primarily selected by the long-range interactions, as the nearest neighbour interactions are close to the degenerate point between zigzag and bulk ferromagnetic order. Following Ref. \onlinecite{radu2016new}, the magnetic configuration can be written in terms of the two-site basis shown in Fig. \ref{fig-linet}(c).:
\begin{align}
\mathbf{S}_1(\mathbf{r}) = \sum_{\mathbf{k}} (\mathbf{F}_{\mathbf{k}}+\mathbf{A}_{\mathbf{k}})e^{-i\mathbf{k}\cdot\mathbf{r}}\\
\mathbf{S}_2(\mathbf{r}) = \sum_{\mathbf{k}} (\mathbf{F}_{\mathbf{k}}-\mathbf{A}_{\mathbf{k}})e^{-i\mathbf{k}\cdot\mathbf{r}}
\end{align}
We assume that a particular order is described by nonzero Fourier components at a single $\mathbf{q}$ vector $\mathbf{F}_\mathbf{q} = \mathbf{F}_{-\mathbf{q}}^*$, $\mathbf{A}_\mathbf{q} = \mathbf{A}_{-\mathbf{q}}^*$. The cartesian components of such order parameters are written in the crystallographic $(\hat{a},\hat{b},\hat{c}^*)$ coordinates. For large $A_1<0, B_1>0$, the classical ground state of model (\ref{eqn-35}) is bulk ferromagnetic order defined by $\mathbf{q}=0$, $\mathbf{F}_\mathbf{q}=(F_a,F_b,0)$, i.e. the moment is confined to the $ab$-plane. For large $J_3>0$ the zigzag state is preferred; for zigzag chains running along $a$, the ordered moment is confined to the $ac^*$-plane, as in Na$_2$IrO$_3$ and $\alpha$-RuCl$_3$. In this case, the propagation vector within the plane is $\mathbf{q}=(0.5,0)$ with respect to the coordinates $(2\pi\hat{a}/a,2\pi\hat{b}/b)$ and $\mathbf{A}_\mathbf{q}=(A_a,0,A_{c^*})$. Conventional Neel order is defined by $\mathbf{q}=0$ and $\mathbf{A}_\mathbf{q}=(0,0,A_{c^*})$.
%, while the stripy state has $\mathbf{q}=(\pi,0,0)$ and $\mathbf{F}_\mathbf{q} = (A_a,0,-A_{c^*})$.

For the initial set of parameters, the classical ground state of model (\ref{eqn-35}) is zigzag, as a result of the large $J_3$. However, there exists a nearby counterrotating incommensurate state for finite $A_2,B_2$ with $\mathbf{q}=(q,0)$ within the $ab$-plane that appears at intermediate $J_3$, as shown in Fig. \ref{fig-liphase}(b). This coplanar spiral phase has the same components as the experimental order parameter $\mathbf{F}_\mathbf{q}=(0,F_b,0), \mathbf{A}_\mathbf{q}=(-iA_a,0,-iA_{c^*})$,\cite{radu2016new} and is adiabatically connected to the incommensurate states identified in the pure nearest neighbour $(J_1,K_1,\Gamma_1)$ model.\cite{rau2014generic} It emerges naturally as an energetic compromise between the nearby ferromagnetic, Neel, and zigzag states. The ordering vector $q$ and ratio of $A_a/A_{c^*}$ are determined by $B_1,A_2,B_2$, and $D$, while $A_1$ and $J_3$ control only the relative energies of the neighbouring states. The experimental values $q_{\rm exp} = 0.32$ and $A_a/A_{c^*} = 0.16$ are obtained for modest second neighbour interactions:
\begin{align}
\frac{B_2}{B_1} \sim 0.3 -1.3\frac{D}{B_1} \ \ \ , \ \ \ \frac{A_2}{B_1} \sim -0.14 +0.5\frac{D}{B_1}
\end{align}
For $D=0$, for example, this corresponds to $B_2>0,A_2<0$, and $B_2/B_1=0.3$ and $A_2/B_2=0.5$, in essential agreement with the computed bond-averaged values. We show in Fig. \ref{fig-linet}(c), the classical ground state for $D=0$ and typical interactions satisfying these criteria: $(A_1,B_1,A_2,B_2,J_3) \sim (-10.7, +24, -3.3,+7.0,+3.6) $ meV. These values differ from the bond-averaged values of the relaxed structure only by an enhanced $K_1=-\Gamma_1=-B_1/2\sim -12$ meV and somewhat reduced $J_3$. The possible effects of the second-neighbour DM terms also deserve serious consideration, as these terms certainly stabilize spiral order. Even a small magnitude $D/K_1 \sim 0.15$, is sufficient to significantly enhance the stability region of the incommensurate phase, as shown in Fig. \ref{fig-liphase}(c). This result can be understood as follows: the honeycomb lattice in $C2/m$ symmetry is bipartite, with sublattices $\mathbf{S}_1$ and $\mathbf{S}_2$ related by inversion. Second neighbour interactions always couple sites belonging to the same sublattice. All such sites within a sublattice are related by translation, implying the $\mathbf{D}_2$ vectors of a given bond-type are {\it uniform} within a given sublattice, as shown in Fig. \ref{fig-linet}(a,b). For this reason, the DM-interaction uniquely promotes spiral states of opposite chirality (i.e. counter rotating) on each sublattice. For the sign and orientation of the computed $\mathbf{D}_2$ vectors, the experimental $(-iA_a, F_b, -iA_{c^*})$ state is stabilized. An example parameter set consistent with the experimental magnetic structure for $D\neq 0$ is: $(A_1,B_1,A_2,B_2,D,J_3) \sim (-8.9, +20, -2.0,+3.9,+1.5,+3.0) $ meV. We note that these last suggested interactions correctly reproduce the enhancement of ferromagnetic terms in $\alpha$-Li$_2$IrO$_3$ compared with Na$_2$IrO$_3$, although a predicted ferromagnetic $\Theta_{iso}>0$ may be incompatible with the experimental observation of $\Theta_{iso}\sim -33$ K.\cite{singh2012relevance} Further refinement of the crystal structure would allow for reevaluation of this result. We also predict a ferromagnetic $\Theta_b\gtrsim\Theta_a>0$, and antiferromagnetic $\Theta_{c^*} < 0$ for $\alpha$-Li$_2$IrO$_3$, which suggest an opposite anisotropy in $\chi$ as observed in Na$_2$IrO$_3$. This result could be confirmed in future single-crystal studies.

Taken together, these results emphasize that the presence of modest long-range $K_2,\Gamma_2,\mathbf{D}_2$ interactions is sufficient to obtain the experimental state of $\alpha$-Li$_2$IrO$_3$ without needing additional bond-anisotropy. However, it is important to emphasize that the computed interactions for the reported experimental structure of $\alpha$-Li$_2$IrO$_3$ also suggest that a bond dependence of the interactions at least on the scale suggested in Ref. \onlinecite{Kimchi:2015ky} is very reasonable. Given the potentially large number of symmetry inequivalent magnetic interactions in these materials, it is difficult to discuss the specific details of all long-range and bond-anisotropic terms, and it should be recognized that {\it all} such interactions can be relevant in the real materials. In a broader context, the results of this section suggest that long-range interactions may also play a significant role in stabilizing the incommensurate order in the 3D $\beta$- and $\gamma$-Li$_2$IrO$_3$.\cite{biffin2014noncoplanar,biffin2014unconventional}

\section{Discussion: Realization of the Spin Liquid in Real Materials}
\label{sec-5}
\begin{figure}[t]
\includegraphics[width=\linewidth]{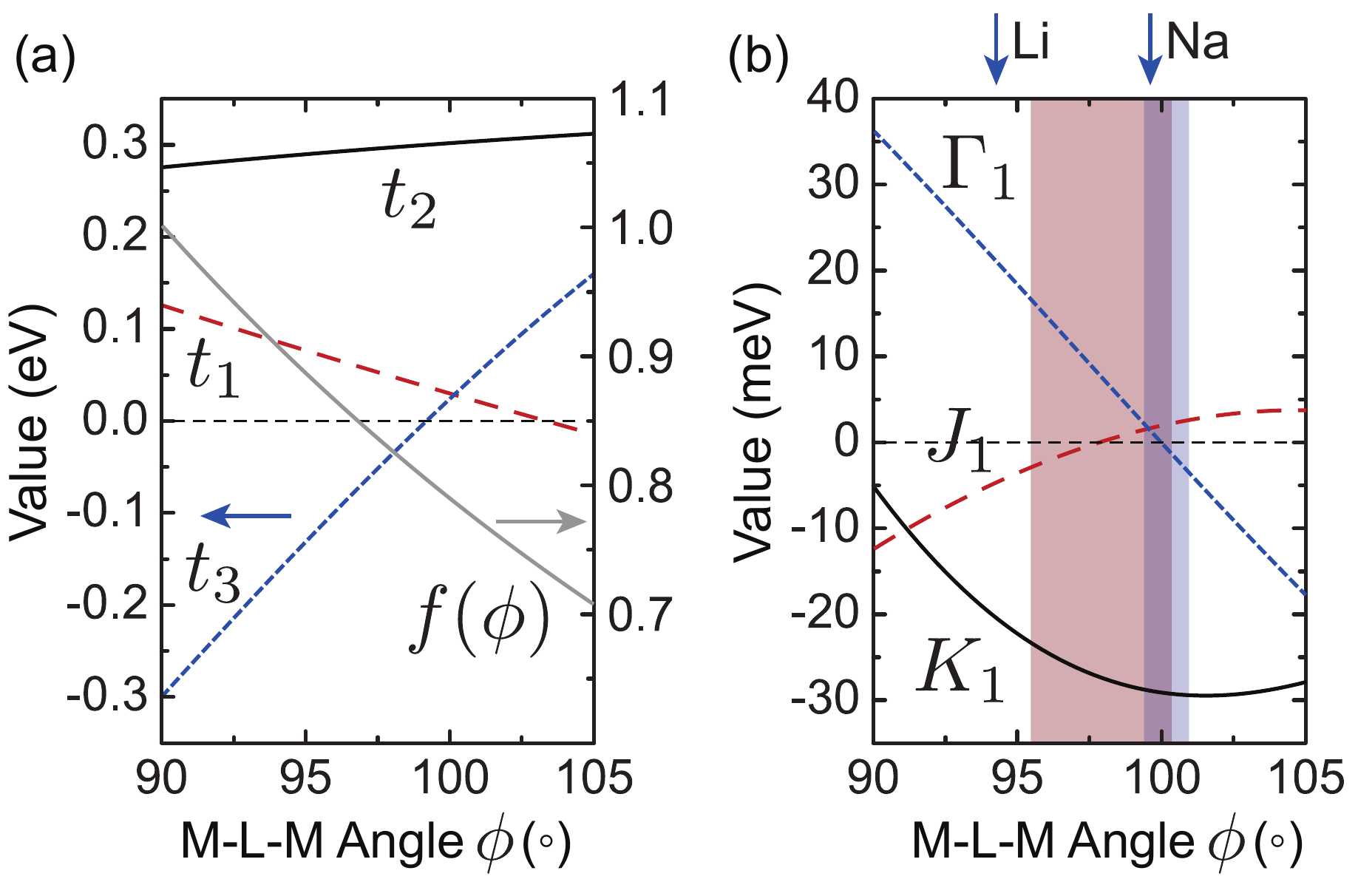}
\caption{\label{fig-angfig} (a) Schematic dependence of hopping integrals $t_{1-3}$ for $A$$_2$IrO$_3$ and empirical damping factor $f(\phi)<1$ (grey line) on metal-ligand-metal bond angle $\phi$; $t_4$ is assumed to be zero. (b) Resulting magnetic interactions obtained using the ``Exact'' 2OPT expressions of section \ref{sec_2OPT}. The red shaded region indicates the area where $|J_1/K_1|<0.1$, while the blue region denotes $|\Gamma_1/K_1|<0.1$. For $\alpha$-Li$_2$IrO$_3$, $\phi\sim 95^\circ$, while Na$_2$IrO$_3$ falls nearly in ideal region $\phi \sim 100^\circ$.}
\end{figure}

In this section, we consider the challenges for designing real materials in the Kitaev spin-liquid phase. It is expected, based on the computed phase diagrams in this and previous works, that the spin-liquid occupies a very small region of parameter space, complicating the search for real materials exhibiting this phase. For example, for the pure Kitaev-Heisenberg $J_1>0,K_1<0$ model, the spin-liquid is thought to require $\alpha=K_1/(K_1-2J_1)\gtrsim 0.7-0.8$ for a variety of lattices.\cite{Chaloupka:2010gi,Lee:2014ib} The effects of further neighbour Heisenberg coupling are mixed; while generic $J_2,J_3$ tend to lift the classical degeneracy and promote order, long-range terms that frustrate the classical order may also open extended spin-liquid regions adiabatically connected to the Kitaev state.\cite{Nishimoto:2014vw,Katukuri:2014iq,singh2012relevance} This effect is seen in Fig. \ref{fig-Namag}. However, when interactions are unfrustrated, we generally expect order to appear unless $|J_3/K_1|\lesssim 0.1$. Similarly, off-diagonal interactions $\Gamma_1,\Gamma_1^\prime$ are generally detrimental to the spin-liquid, with results of exact diagonalization suggesting e.g. $\Gamma_1/K_1 \lesssim 0.1$ is required.\cite{rau2014generic,Rau:2014wc} For the purpose of discussion, we assume these rough boundaries to be accurate, and consider how they might be met in real materials. We approximate the hopping integrals via Slater-Koster parameters\cite{slater1954simplified} for direct metal-metal or metal-ligand hopping through various symmetry channels (i.e. $t_{dd\sigma},t_{dd\pi},t_{pd\pi}$), as per Ref. \onlinecite{rau2014generic}.\footnote{Note that the expression for $t_2$ in the supplemental of Ref. \onlinecite{rau2014generic} appears with different sign conventions.} Given the additional approximations $ t_{dd\delta}=t_4=0$, we have:
\begin{align}\label{eqn-hop1}
t_1 \sim & \  \frac{1}{2}t_{dd\pi}+ \frac{t_{pd\pi}^2}{\Delta_{pd}}\cos \phi \\
t_2 \sim & \ - \frac{1}{2}t_{dd\pi}+ \frac{t_{pd\pi}^2}{\Delta_{pd}}  \\  \label{eqn-hop2}
t_3 \sim & \ \frac{3}{4}t_{dd\sigma} + \frac{(t_{pd\pi} -\sqrt{3}t_{pd\sigma})^2}{8\Delta_{pd}} \cos 3\phi \\ \nonumber & \ +\frac{(\sqrt{3}t_{pd\pi}+9t_{pd\sigma})(\sqrt{3}t_{pd\pi}+t_{pd\sigma})}{8\Delta_{pd}}\cos \phi
\end{align}
where $\phi$ is the metal-ligand-metal bond angle, and $\Delta_{pd}$ gives the charge transfer energy between the $t_{2g}$ orbitals and bridging chalcogen or halogen $p$-orbitals. For the $A$$_2$IrO$_3$ materials, values consistent with the computed hopping integrals are obtained for:
\begin{align}
t_{dd\pi} \sim 0.25 f(\phi) \text{ eV}\ \ \ ,& \  \ \ \ t_{dd\sigma} \sim -0.4 f(\phi) \text{ eV}  \\
\frac{t_{pd\pi}^2}{\Delta_{pd}} \sim 0.4 \text{ eV} \ \ \ ,& \ \ \ \ \frac{t_{pd\sigma}^2}{\Delta_{pd}} \sim 0.5 \text{ eV}
\end{align}
where $f(\phi)$ is an approximate exponential damping factor (Fig. \ref{fig-angfig}(a)) intended to capture the suppression of direct metal-metal hopping with increasing Ir-Ir bond distance. Here we have assumed that the Ir-O distances are constant, and that the $p_x$, $p_y$ and $p_z$ orbitals of the oxygen ligands are roughly degenerate. For $\phi=90^\circ$, ligand mediated hopping contributes only to $t_2$, while $t_1$, $t_3$ are dominated by direct metal-metal hopping. For $\phi>90^\circ$, the distortion allows ligand-mediated contributions to $t_1$ and $t_3$, which drives both terms to nearly zero around $\phi \sim 100^\circ$, as shown in Fig. \ref{fig-angfig}(a). We now discuss the consequences for the magnetic interactions.

1. {\it Heisenberg coupling:} We recall that, at $\mathcal{O}(t^2)$, for $\Delta_n=0$, the exact nearest neighbour interactions depend on constants $\mathbb{A},\mathbb{B}(U,J_{\rm H},\lambda)$. The values of these constants were estimated above for real materials; for $5d$ Ir$^{4+}$ ions $
 \mathbb{A}_{5d}\sim 0.9 \text{ eV}^{-1} , \mathbb{B}_{5d}\sim 0.04 \text{ eV}^{-1}$
while for $4d$ Ru$^{3+}$ ions (as in $\alpha$-RuCl$_3$), $ \mathbb{A}_{4d}\sim 0.6 \text{ eV}^{-1} ,\mathbb{B}_{4d}\sim 0.05 \text{ eV}^{-1}$.
Taken together, these values suggest that the charge gap $(\sim \mathbb{A}^{-1})$ and natural scale for the magnetic interactions are not strongly dependent on the choice of $4d^5$ or $5d^5$ ion. The finding $\mathbb{A}\gg\mathbb{B}$ 
which implies an arbitrary set of hopping terms would give $J_1 > 0$ and $J_1 \gg K_1,\Gamma_1,\Gamma^\prime_1$ far away from the Kitaev limit. Indeed, the natural energy scale for the isotropic exchange is much greater than the anisotropic terms,\cite{jackeli2009mott} so suppression of $J_1$ should generally require fine tuning of the hopping integrals $t_{1},t_3,t_4$. However, the finding $|J_1|<|K_1|$ in all materials studied in this work suggests that the suppression of $J_1$ is relatively robust in real systems. This is due to the typical relationship between hopping integrals $t_1$ and $t_3$, which ensure that $\mathbb{A}(2t_1+t_3) \sim 0$ over a large region of $\phi$. For this reason, the suggested $\alpha \gtrsim 0.8$ requirement for the spin liquid is actually satisfied for a wide region $95^\circ \lesssim \phi \lesssim 100^\circ$, as indicated by the red shaded region of Fig. \ref{fig-angfig}(b). The main synthetic challenge for approaching the Kitaev limit is therefore controlling other terms in the Hamiltonian such as the off-diagonal $\Gamma_1,\Gamma_1^\prime$ and longer-range interactions.

2. {\it Off-diagonal terms $\Gamma_1$, $\Gamma_1^\prime$:} Inspection of Eq. (\ref{eqn-J1})$-$(\ref{eqn-Gp1}) reveals that $t_2\gg t_1,t_3,t_4$ is the {\it only} limit where the pure Kitaev model is realized at $\mathcal{O}(t^2)$. This is true because all anisotropic terms scale with the same constant $\propto \mathbb{B}$, so that large off-diagonal terms always appear at the same order as $K_1$ for finite $t_1,t_3,t_4$. Given the above approximations, we find only a narrow region near $\phi \sim 100^\circ$ where $|\Gamma_1/K_1|<0.1$, implying a significantly more severe restriction on $t_1,t_3$ than the suppression of $J_1$. Surprisingly, Na$_2$IrO$_3$ appears to be very close to this ideal region as a direct result of trigonal distortion.  For this reason, distortions are initially helpful for realizing the spin-liquid, contrary to initial assumptions. However, large distortions also imply large off-diagonal hopping $t_4$ and crystal field splitting $\Delta_n$, which have not been considered here. These provide alternative contributions to $\Gamma_1,\Gamma_1^\prime$.\cite{Rau:2014wc,khaliullin2005orbital,PhysRevB.92.024413} Finding an optimal balance between these effects is clearly required for stabilizing the spin-liquid. 

It may be possible, in principle, to shift or enhance the stability region of the spin-liquid through expansion of the lattice, in order to decrease the overall scale of direct metal-metal hopping and therefore $t_1,t_3$. In theory, this may be accomplished in thin-films or heterostructures of 2D honeycomb layers with a suitable substrate, but it is important not to introduce other symmetry-lowering distortions or surface potentials. For $\phi\neq 90^\circ$, ligand mediated hopping contributes to both $t_2$ and $t_1,t_3$, so that large distortions $\Delta\phi>10^\circ$ are unfavourable in combination with lattice expansion.  That is, the ideal region is shifted toward $\phi=90^\circ$ as direct metal-metal hopping is decreased. An alternative strategy for decreasing $|t_1/t_2|$ in the bulk could be the incorporation of heavier ligands such as S or Se in place of O or Br and I in place of Cl. This should suppress $t_{dd\sigma}$ and $t_{dd\pi}$ by elongating the metal-metal distances, while enhancing $t_2$ via increased covalency (i.e. smaller $\Delta_{pd}$). 
The obvious challenge is to find insulating materials with the correct lattice geometry and metal valency. 
 To the best of our knowledge, there are no edge-sharing honeycomb $d^5$ materials incorporating heavy Br, I, S or Se ligands so far reported. The natural downside of incorporating heavier ligands is that the enhancement of $t_2/U$ and long-range hopping may also result in stronger long-range interactions, or even an itinerant state.

3. {\it Long-range interactions:} 
For the $d^5$ spin-orbit assisted Mott insulators on edge-sharing lattices, suppression of long-range interactions is complicated by two effects. The first is that significant second and third neighbour hopping terms always arise from M-L-L-M hopping processes through close L-L contacts. These produce large long-range interactions at relatively low orders in perturbation theory $J_2,J_3 \sim t^3/U_\text{eff}^2$. In principle, these terms may still be suppressed for large $U_\text{eff}$. However, this observation is not synthetically useful because of a second effect. The energy scale for virtual processes $U_\text{eff}\sim \mathbb{A}^{-1}$ depends in a complementary way on $U$, $\lambda$ and $J_{\rm H}$, ensuring roughly similar charge gaps for $4d^5$ and $5d^5$ materials. In the lighter $4d$ systems, the larger $U$ is partially offset by a larger $J_{\rm H}$ and smaller $\lambda$. This point can be seen from analysis of the optical excitation spectra of Na$_2$IrO$_3$ and $\alpha$-RuCl$_3$.\cite{Sears:2015ku, Comin:2012jz,Plumb:2014hh,Yingli2015,Sandilands:2015vk} As with Ref. \onlinecite{Kim:2012gu,Kim:2014kn} we find, in our ED calculations, that the lowest energy excitations are essentially local on-site $j_{\text{eff}}=1/2 \ \rightarrow \ j_{\text{eff}}=3/2$ transitions, which are expected to be weakly absorbing. For $\alpha$-RuCl$_3$, these are predicted in the range $\Delta E = 3\lambda/2 \sim 0.18-0.28$ eV, and naturally explain the sharp and weak transitions observed experimentally in this frequency region. In the ED results, charge carrying intersite excitations were found at higher energy, in the range $\Delta E \sim 1.4-1.7$ eV $\sim \mathbb{A}^{-1}$, which corresponds roughly with the first intense peak in the experimental $\epsilon_2(\omega)$ at $\omega \sim U_\text{eff} \sim 1.2$ eV.\cite{Plumb:2014hh,Sandilands:2015vk} For Na$_2$IrO$_3$, the enhancement of spin-orbit coupling shifts the local excitations to $\Delta E \sim 0.35-0.7$ eV $ \sim 3\lambda /2$, which results in significant mixing with the higher energy intersite excitations, and therefore a soft charge gap. While the scale of $U_\text{eff}$ is therefore obscured in $\sigma_1 (\omega)$, one can expect $0.7 \lesssim \mathbb{A}^{-1} \lesssim 1.5$ for Na$_2$IrO$_3$, in the same range as observed for $\alpha$-RuCl$_3$. Given the similar $U_\text{eff}$ in both $4d$ and $5d$ systems, suppression of long-range terms through incorporation of lighter elements appears unlikely. While we have not presented an analytical treatment of the high order contributions to $\mathbf{J}_2,\mathbf{J}_3$, the ED calculations presented in the previous section imply that third neighbour Heisenberg coupling $J_3$ is very robust for the $d^5$ honeycomb materials. This term explains the prevalence of zigzag order in both Na$_2$IrO$_3$ and $\alpha$-RuCl$_3$. Given that $J_3$ is not frustrated, it is strongly detrimental for realizing the spin-liquid state. In principle, lattice expansion could also suppress $J_3$ in the honeycomb materials if $t/U$ could be decreased. This interaction could also be frustrated by large second neighbour terms, as in $\alpha$-Li$_2$IrO$_3$. Alternately, we note that some of the perturbative contributions to $J_3$ shown in Fig. \ref{fig-highorder} are absent in the 3D $\beta$- and $\gamma$-Li$_2$IrO$_3$, as the closed loops of these lattices may be larger than six sites.\cite{modic2014realization, takayama2015hyperhoneycomb} For this reason, long-range interactions should be partially suppressed in the 3D systems. However, it is not yet clear whether the $|J_3/K_1|\lesssim 0.1$ requirement can be met in real materials.

\section{Summary and Conclusions}
\label{sec-6}

 In the $C2/m$ honeycomb Ir$^{4+}$ and Ru$^{3+}$ systems, complexity arises from a combination of (i) competing Coulomb, hopping, and spin-orbit energy scales, (ii) relatively low symmetry, (iii) suppression of dominant magnetic couplings, (iii) strongly anisotropic interactions, and (v) significant long-range terms. The details of the interactions in the real materials and their relationship to the experimental properties have therefore
 been intensively debated in the literature.
  In this work, we have addressed this debate by employing nonperturbative exact diagonalization methods that treat interactions at all scales on the same level, and therefore allow estimation of all parameters. The salient conclusions are as follows:
 
 (a) The observed zigzag order in Na$_2$IrO$_3$ and $\alpha$-RuCl$_3$ is explained naturally in terms of a large third neighbour Heisenberg coupling $J_3$ that emerges as a dominant term at high orders in perturbation theory, and was therefore neglected or underestimated in most of previous studies. 
 
 (b) Off-diagonal couplings $\Gamma_1\approx -K_1$ dominate the nearest neighbour magnetic interactions in $\alpha$-RuCl$_3$ and $\alpha$-Li$_2$IrO$_3$ as a result of direct metal-metal (M-M) hopping. These terms can be suppressed by increasing the M-M bond distance, through the distortion of the local ML$_6$ octahedra to provide M-L-M bond angles $\phi\sim 100^\circ$. In the known materials, this ideal region is most closely approached by Na$_2$IrO$_3$, which is therefore the closest to the Kitaev limit $K_1\gg J_1,\Gamma_1$ at the nearest neighbour level. Due to the effects of direct metal-metal hopping, the ideal materials will therefore not be found with $\phi = 90^\circ$, as originally proposed.
 
(c) Although the Kitaev spin liquid is thought to be stable for a finite region of magnetic parameters, the design limitations in real materials are highly restrictive due to a large sensitivity of the interactions to structural details. This sensitivity allows for large variations in the magnitude of interactions along the different non-equivalent bonds, which typically lifts the classical degeneracy. The ideal region where the Kitaev interaction is dominant is likely confined to a small width of M-L-M bond angle $\Delta\phi \lesssim 1^\circ$, which may be difficult to satisfy in real materials simultaneously for all non-equivalent bonds.

(d) Given that Na$_2$IrO$_3$ was found to lie very close to the ideal region where $K_1\gg \Gamma_1,J_1$, the most significant interaction preventing realization of the spin-liquid state in real materials is considered to be the unfrustrated long-range $J_3$ term. In the $d^5$ materials, the complementary nature of spin-orbit coupling, and Coulomb repulsion in establishing the charge gap makes $J_3$ largely insensitive to choice of magnetic ions or other structural details. This observation seriously complicates any synthetic strategies aimed at reducing long-range couplings in edge-sharing octahedral systems. 
 
 (e) For $\alpha$-Li$_2$IrO$_3$, the computed interactions suggest the possibility of large bond-anisotropy and significant terms at first, second, and third neighbour. While several model Hamiltonians have been considered for the $\alpha$-,$\beta$-, and $\gamma$-phase materials, the true interactions are likely considerably more complicated. We have shown, in particular, that a combination of $K_2,\Gamma_2$, and second neighbour DM-interaction $\mathbf{D}_2$ may explain the observed order. The complexity of the interactions may be even greater for the lower symmetry $\beta$-, and $\gamma$-Li$_2$IrO$_3$, where Dzyaloshinskii-Moriya interactions are allowed even for certain first neighbour bonds. It remains to be determined which models can be effectively related to the real materials, but purely nearest neighbour models are likely unrealistic.

Given these observations, realization of the Kitaev spin liquid as a ground state in edge-sharing $d^5$ materials appears to represent a very significant synthetic challenge. However, given the highly complex phase diagrams, and possibility of many points of classical degeneracy within the expanded range of interactions, these systems are likely to host other exotic phases, and phase transitions. Furthermore, when probed at high energies or temperatures $T>T_N$, the combined fluctuations associated with all nearby orders may give rise to novel thermodynamic or spectral properties.\cite{Nasu2016a,Nasu2016b,nasu2014vaporization,Yamaji:2016ws} Given the potential for complex interactions, future studies of such systems will benefit from comprehensive and non-perturbative {\it ab-initio} estimates of all relevant interactions. 

\section{Acknowledgements}
The authors would like to acknowledge useful discussions with Kira Riedl, Radu Coldea
and Giniyat Khaliullin. S. M. W. acknowledges support through an NSERC Canada Postdoctoral Fellowship. Y.L. acknowledges support through a
China Scholarship Council (CSC) Fellowship.  H.O.J and
R.V. acknowledge support by the Deutsche Forschungsgemeinschaft through grant SFB/TR 49.

\appendix

\section{Cluster Exact Diagonalization Method}
\label{app_ED}

As in conventional perturbation theory we divide the total Hamiltonian into $\mathcal{H}_0 = \mathcal{H}_{\rm SO}+\mathcal{H}_U$ and $\mathcal{H}_1 =  \mathcal{H}_{\rm hop}+\mathcal{H}_{\rm CF}$.
The Hilbert space is divided according to the energy with respect to $\mathcal{H}_0$. States below some energy cutoff of order $\mathcal{O}(U, \lambda)$ are those in the so-called lower energy subspace $\{|n_l\rangle\}$, and represent pseudospin states to be included in the Hilbert space of the final $\mathcal{H}_{\rm spin}$. In this case, $\{|n_l\rangle\}$ contains the lowest Kramers' doublet on every site; there are $N_l=2^p$ such states, where $p$ is the number of sites in the cluster. All other states fall into the high energy subspace $\{|n_h\rangle\}$ and will be effectively integrated out. We first diagonalize $\mathcal{H}_{tot}$ to obtain the exact eigenstates $\{|n\rangle\}$, which are ordered according to their exact energy $E_n$. As in perturbation theory, the goal is to produce an effective $\mathcal{H}_{\rm spin}$ that reproduces the spectrum $E_n$ for only the $N_l$ lowest states, but is written in terms of the spin basis $\{|n_l\rangle\}$.

We first introduce an intermediate basis $\{|n^\prime\rangle\}$ which is obtained by projecting the lowest $N_l$ exact eigenstates onto the low energy subspace:
\begin{align}
|n^\prime\rangle = \sum_{n_l}^{N_l} |n_l\rangle \langle n_l|n\rangle\equiv \mathbf{R}|n\rangle \ \ , \ \ n,n^\prime = 1 \ ... \ N_l
\end{align}
Since the lower Hilbert space does not form a complete set of states for $\mathcal{H}_{tot}$, the intermediate basis $\{|n^\prime\rangle\}$ is not generally orthonormal despite spanning the lower Hilbert space, i.e. $\mathbf{R}$ is not a unitary transformation. We therefore perform a symmetric orthogonalization of the intermediate basis in terms of the overlap matrix $\mathbf{S}$, in order to define a total unitary transformation:
\begin{align}
\mathbf{U} = \mathbf{R}\mathbf{S}^{-1/2} \ \ \ , \ \ \ \mathbf{S}_{mn} = \langle m^\prime | n^\prime \rangle
\end{align}
The final spin Hamiltonian is then given by:
\begin{align}
\mathcal{H}_{\rm spin} =\mathbf{U}^\dagger\left( \sum_n^{N_l}E_n |n\rangle \langle n | \right) \mathbf{U}
\end{align}
This method is formally similar to the Contractor Renormalization Group (CORE) method at one step,\cite{morningstar1994contractor,capponi2004numerical} but symmetric orthogonalization is employed rather than the Gram-Schmidt procedure. It is easy to show that the resulting $\mathcal{H}_{\rm spin}$ respects all symmetries of the chosen cluster and reproduces the exact low-energy spectrum independent of the choice of $\{|n_l\rangle\}$. The only condition is that $\mathbf{S}$ is nonsingular, which is easily satisfied in practice. For highly symmetric clusters, these conditions are sufficient to constrain the resulting $\mathcal{H}_{\rm spin}$ to be nearly independent of choice of projection basis $\{|n_l\rangle\}$. In all cases, the the method maximizes the overlap between the eigenstates of $\mathcal{H}_{\rm spin}$ and the low energy eigenstates of $\mathcal{H}_{tot}$, which removes most of the ambiguities arising from the choice of projection basis. 
In practice, we always choose $\{|n_l\rangle\}$ to be pure $j_{\text{eff}}=1/2$ states, for which we typically find the eigenvalues of the overlap matrix $\mathbf{S}$ to be large, indicating strong overlap with the exact low-energy eigenstates. Even when the projection basis is chosen very poorly (overlap $<$ 0.5), the correct $\mathcal{H}_{\rm spin}$ is typically restored by the orthogonalization step, as long as $\{|n_l\rangle\}$ and $\{|n\rangle\}$ are adiabatically connected to one another. For example, the method correctly reproduces the behaviour expected at both weak\cite{Rau:2014wc} and strong\cite{Bhattacharjee:2012hk} trigonal distortion despite poor overlap of the $j_{\text{eff}}=1/2$ states in the latter case. Errors generated by ambiguity of the projection basis are estimated as $\sim \pm 1$ meV.
\section{Further Neighbour Hopping Integrals}
\label{app_hop}

In the $C2/m$ materials, the presence of distortions implies that metal-ligand-metal bond angles differ from 90$^\circ$, allowing for ambiguity in choice of cubic coordinates $x,y,z$. In this work, coordinates for the projections were defined as in Ref. \onlinecite{Foyevtsova:2013uo}: the local $z$-axis was chosen to be $\hat{z}\perp (\mathbf{a}+\mathbf{c})$, while the remaining axes were uniquely defined by requiring $x,y\perp z$, and both axes make a $45^\circ$ angle with the crystallographic $b$-axis. First neighbour hopping for the experimental structures was given in Section \ref{sec_hop}. Further neighbour hopping is given in Tables \ref{table_further} and \ref{table_further2}.
\begin{figure*}[t]
\includegraphics[width=0.85\linewidth]{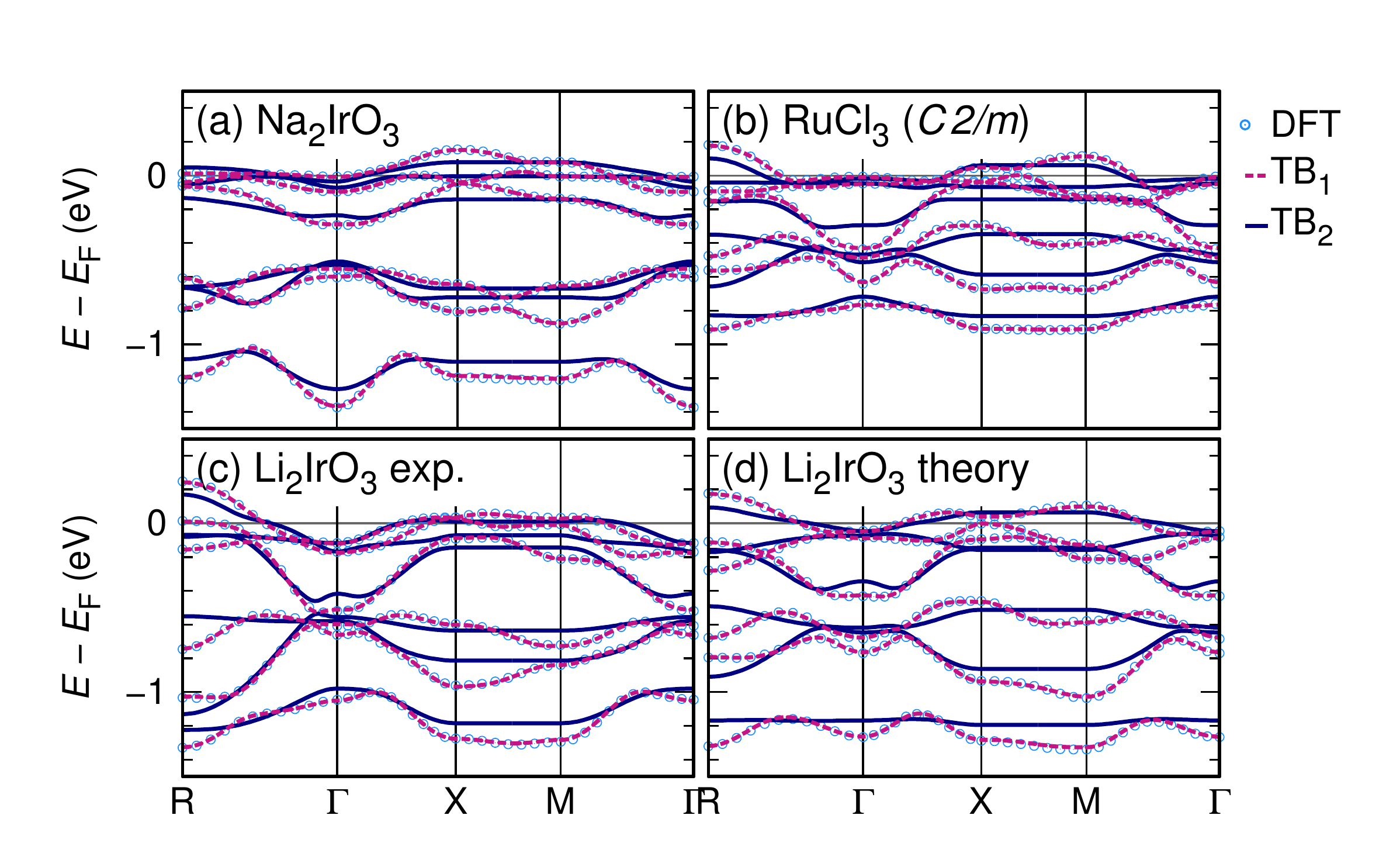}
\caption{\label{fig-bands} Comparison of DFT computed bandstructures (blue circles) and tight-binding
interpolations employing hopping integrals up to 16 \AA  \ (TB$_1$, dashed pink line), and 3rd neighbours (TB$_2$, solid blue line).}
\end{figure*}

\begin{table}[t]
\caption {\label{table_further}Hopping parameters for second nearest neighbours (meV).}
\centering\def\arraystretch{1.0}\small
\begin{ruledtabular}
\begin{tabular}{ccccccc}
Bond&&{\nairo}&\multicolumn{2}{c}{$\alpha$-Li$_2$IrO$_3$}&$\alpha$-{\ruclc}  \\
&&&Exp.&Theory&$C2/m$\\
\hline
X$_2$:& $xz$ $\rightarrow$ $xz$ & -0.8   & -0.6   & -3.7   & -4.5  \\
& $yz$ $\rightarrow$ $yz$ & -1.6   & 3.6    & 0.2    & -0.4      \\
& $xy$ $\rightarrow$ $xy$ & -3.6   & 1.2    & -2.3   & -3.2 \\
& $xy$ $\rightarrow$ $xz$ & -75.8  & -56.9  & -70.5  & -59.1  \\
& $xz$ $\rightarrow$ $xy$ & -36.4  & -23.8  & -38.6  & -24.3   \\
& $xy$ $\rightarrow$ $yz$ & 12.7   & 15.2   & 11.0   & 8.3   \\
& $yz$ $\rightarrow$ $xy$ & -21.3  & -10.4  & -10.2  & 1.3     \\
& $xz$ $\rightarrow$ $yz$ & -18.4  & -16.4  & -11.6  & -1.2      \\
& $yz$ $\rightarrow$ $xz$ & 10.3   & 28.9   & 11.8   & 11.8   \\
\hline
Y$_2$:& $yz$ $\rightarrow$ $yz$ & -0.8   & -0.6   & -3.7   & -4.5       \\
& $xz$ $\rightarrow$ $xz$ & -1.6   & 3.6    & 0.2    & -0.4    \\
& $xy$ $\rightarrow$ $xy$ & -3.6   & 1.2    & -2.3   & -3.2     \\
& $yz$ $\rightarrow$ $xy$ & -75.8  & -56.9  & -70.5  & -59.1    \\
& $xy$ $\rightarrow$ $yz$ & -36.4 & -23.8  & -38.6  & -24.3    \\
& $xz$ $\rightarrow$ $xy$ & 12.7   & 15.2   & 11.0   & 8.3      \\
& $xy$ $\rightarrow$ $xz$ & -21.3  & -10.4  & -10.2  & 1.3      \\
& $xz$ $\rightarrow$ $yz$ & -18.4  & -16.4  & -11.6  & -1.2     \\
& $yz$ $\rightarrow$ $xz$ & 10.3   & 28.9   & 11.8   & 11.8    \\
\hline
Z$_2$:& $xy$ $\rightarrow$ $xy$ & -1.5   & 1.0    & 0.6    & -0.4    \\
& $xz$ $\rightarrow$ $xz$ & -1.6   & -1.2   & -2.9   & -4.7      \\
& $yz$ $\rightarrow$ $yz$ & -1.6   & -1.2   & -2.9   & -4.7     \\
& $xz$ $\rightarrow$ $yz$ & -77.0  & -56.7  & -73.2  & -60.7   \\
& $yz$ $\rightarrow$ $xz$ & -30.3  & -51.4  & -39.6  & -23.9 \\
& $xy$ $\rightarrow$ $xz$, $yz$ $\rightarrow$ $xy$ & -18.8  & -12.5  & -11.3  & -1.7     \\
& $xy$ $\rightarrow$ $yz$, $xz$ $\rightarrow$ $xy$ & 9.4    & 5.3    & 11.6   & 11.6      \\
\end{tabular}
\end{ruledtabular}
\end{table}
\begin{table}[h]
\caption {\label{table_further2}Hopping parameters for third nearest neighbours (meV).}
\centering\def\arraystretch{1.0}\small
\begin{ruledtabular}
\begin{tabular}{llrrrrr}
Bond&&{\nairo}&\multicolumn{2}{c}{$\alpha$-Li$_2$IrO$_3$}&$\alpha$-{\ruclc}  \\
&&&Exp.&Theory&$C2/m$\\
\hline
X$_3$:& $yz$ $\rightarrow$ $yz$ & -35.3 & -40.0  & -33.0  & -41.4   \\
&  $xy$ $\rightarrow$ $xy$ & -8.5  & -3.5  & -5.8   & -7.5     \\
&  $xz$ $\rightarrow$ $xz$ & -8.2  & -11.2 & -6.8   & -7.9   \\
& $xy$ $\rightarrow$ $xz$, $xz$ $\rightarrow$ $xy$ & -12.7 & -13.0  & -13.4  & -7.8   \\
& $xy$ $\rightarrow$ $yz$, $yz$ $\rightarrow$ $xy$  & 17.0  & 9.5    & 15.3   & 10.7    \\
& $xz$ $\rightarrow$ $yz$, $yz$ $\rightarrow$ $xz$ & 14.9  & 13.1   & 16.3   & 12.7   \\
\hline
Y$_3$:& $xz$ $\rightarrow$ $xz$ & -35.3 & -40.0  & -33.0  & -41.4   \\
&  $xy$ $\rightarrow$ $xy$ & -8.5  & -3.5  & -5.8   & -7.5   \\
&  $yz$ $\rightarrow$ $yz$ & -8.2  & -11.2 & -6.8   & -7.9     \\
& $yz$ $\rightarrow$ $xy$, $xy$ $\rightarrow$ $yz$ & -12.7 & -13.0  & -13.4  & -7.8     \\
& $xy$ $\rightarrow$ $xz$, $xz$ $\rightarrow$ $xy$ & 17.0  & 9.5    & 15.3   & 10.7     \\
& $yz$ $\rightarrow$ $xz$, $xz$ $\rightarrow$ $yz$ & 14.9  & 13.1   & 16.3   & 12.7     \\
\hline
Z$_3$:& $xy$ $\rightarrow$ $xy$ & -36.8 & -40.8  & -33.3  & -39.5  \\
&  $xz$ $\rightarrow$ $xz$, $yz$ $\rightarrow$ $yz$ & -9.3  & -8.1  & -6.4   & -8.2    \\
& $xz$ $\rightarrow$ $yz$, $yz$ $\rightarrow$ $xz$ & -13.8 & -13.6  & -13.5  & -7.4   \\
& $xz$ $\rightarrow$ $xy$, $xy$ $\rightarrow$ $xz$ & 16.6  & 15.8   & 16.6   & 11.7    \\
& $yz$ $\rightarrow$ $xy$, $xy$ $\rightarrow$ $yz$ & 16.6  & 15.8   & 16.6   & 11.7     \\
\end{tabular}
\end{ruledtabular}
\end{table}

\section{Complete Expressions for Nearest Neighbour Interactions at $\mathcal{O}(t^2)$ for $\Delta_n=0$}
\label{app_terms}

For arbitrary symmetry, it is conventional to write the spin Hamiltonian as:
\begin{align}
\mathcal{H}_{\rm spin} = \sum_{ij} J_{ij} \  \mathbf{S}_i \cdot \mathbf{S}_j + \mathbf{D}_{ij} \cdot \mathbf{S}_i \times \mathbf{S}_j + \mathbf{S}_i \cdot \mathbf{\Gamma}_{ij} \cdot\mathbf{S}_j
\end{align}
where $\mathbf{D}_{ij}$ is the Dzyaloshinskii-Moriya (DM) vector, and the traceless tensor $\mathbf{\Gamma}_{ij}$ characterizes the pseudo-dipolar interaction. This corresponds to the parameterization:
\begin{align}
\mathbf{J}_{ij}=\left(\begin{array}{ccc} J_{ij}+\Gamma_{ij}^{aa} &D_{ij}^c+ \Gamma_{ij}^{ab} & -D_{ij}^b+\Gamma_{ij}^{ac} \\ -D_{ij}^c+\Gamma_{ij}^{ab} & J_{ij}+\Gamma_{ij}^{bb} & D_{ij}^a+\Gamma_{ij}^{bc} \\ D_{ij}^b+\Gamma_{ij}^{ac}& -D_{ij}^a+\Gamma_{ij}^{bc} & J_{ij}+\Gamma_{ij}^{cc}\end{array}\right)
\end{align}
In the absence of crystal field splitting, the effective spin Hamiltonian obtained at second order in hopping may be written:
\begin{align}
\mathcal{H}_{\rm spin} = \sum_{i,j}\mathbb{P}\mathbf{c}^\dagger_i \mathbf{T}_{ij} \mathbf{c}_j \mathbb{Q}\left(\omega-\mathcal{H}_{0} \right)^{-1}\mathbb{Q}\mathbf{c}^\dagger_j \mathbf{T}_{ji} \mathbf{c}_i\mathbb{P}
\end{align}
where $\mathbb{P}$ is the projection operator into the lower $j_{\text{eff}}=1/2$ subspace, $\mathbb{Q}=1-\mathbb{P}$, and $\mathcal{H}_0 = \mathcal{H}_{\rm SO}+\mathcal{H}_{U}$. For expressions correct to second order, we take $\omega=\langle \mathcal{H}_0\rangle$, the expectation value in the lower subspace. The perturbation theory is most conveniently formulated in terms of the one-particle eigenstates of $\mathcal{H}_{\rm SO}$, labelled $|j,m_j\rangle$. For this reason, we write:
\begin{align}
\mathbf{c}_{i} = \left(\begin{array}{c}\mathbf{c}_{i,\frac{1}{2}} \\ \mathbf{c}_{i,\frac{3}{2}} \end{array} \right)
\end{align}
where:
\begin{align}
\mathbf{c}_{i,\frac{1}{2}} = \left(\begin{array}{c}c_{i,|\frac{1}{2},\frac{1}{2}\rangle} \\ c_{i,|\frac{1}{2},-\frac{1}{2}\rangle} \end{array} \right) \ \ , \ \ \mathbf{c}_{i,\frac{3}{2}} = \left(\begin{array}{c} c_{i,|\frac{3}{2},\frac{3}{2}\rangle} \\ c_{i,|\frac{3}{2},\frac{1}{2}\rangle} \\ c_{i,|\frac{3}{2},-\frac{1}{2}\rangle} \\ c_{i,|\frac{3}{2},-\frac{3}{2}\rangle} \end{array} \right)
\end{align}
Here, $\mathbf{c}_{i,\frac{1}{2}}$ creates a hole in the $j_{\text{eff}}=1/2$ state, and $\mathbf{c}_{i,\frac{3}{2}}$ creates a hole in the $j_{\text{eff}}=3/2$ state. In the absence of crystal field splitting, the zeroth order projection operator into the low energy space is simply:
\begin{align}
\mathbb{P} = \prod_i \mathbf{c}^\dagger_{i,\frac{1}{2}} \mathbf{c}_{i,\frac{1}{2}}
\end{align}
We write the magnetic Hamiltonian in terms of the spin operators in the $j_{\text{eff}}=1/2$ basis:
\begin{align}
\mathbf{S}_i =\frac{1}{2}\mathbf{c}_{i,\frac{1}{2}}^\dagger \vec{\sigma}\mathbf{c}_{i,\frac{1}{2}}\label{eqn-spin}
\end{align}
where $\vec{\sigma}=(\sigma_x,\sigma_y,\sigma_z)$ is the Pauli vector:
\begin{align}
\sigma_x = \left(\begin{array}{cc}0 & 1 \\ 1 & 0 \end{array} \right) ,\sigma_y = \left(\begin{array}{cc}0 & -i \\ i & 0 \end{array} \right),\sigma_z = \left(\begin{array}{cc}1 & 0 \\ 0 & -1 \end{array} \right)
\end{align}
Magnetic interactions are easily evaluated by exact calculation of the propagator $G_j^0$:
\begin{align}
G_j^0=\mathbf{c}_j \mathbb{Q}\left(\omega-\mathcal{H}_{0} \right)^{-1}\mathbb{Q}\mathbf{c}^\dagger_j
\end{align}
acting on a ground state with one hole per site in the $j_\text{eff}=1/2$ state. In matrix form, this can be written:
\begin{align}
G_j^0 = \left(\begin{array}{cc}\mathbb{A}\left(\mathbf{S}_j\cdot\vec{\sigma} - \frac{1}{2}\mathbb{I}_{2\times 2}\right) & 0 \\ 0 & \mathbb{B} \ \mathbf{S}_j\cdot \vec{\tau} - \mathbb{C}\ \mathbb{I}_{4\times 4} \end{array}\right)
\end{align}
where $\mathbb{I}_{n\times n}$ is the $n\times n$ identity matrix, 
and $\vec{\tau}=(\tau_x,\tau_y,\tau_z)$ is the higher dimensional $J=3/2$ equivalent of the Pauli vector:
\begin{align}
\tau_x = \left(\begin{array}{cccc}0 & \sqrt{2} & 0 & 0\\ \sqrt{2} & 0&3&0\\0&3&0&\sqrt{2}\\0&0&\sqrt{2}&0 \end{array} \right)
\end{align}
\begin{align}
 \tau_y =  i\left(\begin{array}{cccc}0 & -\sqrt{2} & 0 & 0\\ \sqrt{2} & 0&-3&0\\0&3&0&-\sqrt{2}\\0&0&\sqrt{2}&0 \end{array} \right)
\end{align}
\begin{align}
\tau_z = \left(\begin{array}{cccc}3 & 0 & 0 & 0\\ 0 & 1&0&0\\0&0&-1&0\\0&0&0&-3 \end{array} \right)
\end{align}
The relevant constants are:
\begin{align}
\mathbb{A} = & \ -\frac{1}{3}\left\{\frac{J_{\rm H} + 3(U+3\lambda)}{6J_{\rm H}^2 - U(U+3\lambda)+J_{\rm H}(U+4\lambda)} \right\}\\
\mathbb{B} = & \ \frac{4}{3}\left\{\frac{(3J_{\rm H}-U-3\lambda)}{(6J_{\rm H}-2U-3\lambda)}\eta\right\}\\
\mathbb{C} =& \  \frac{6}{8}\left\{ \frac{1}{2U-6J_{\rm H}+3\lambda}+\frac{5}{9}\frac{(3U-7J_{\rm H}-9\lambda)}{J_{\rm H}}\eta\right\}\\
\eta= & \ \frac{J_{\rm H}}{6J_{\rm H}^2-J_{\rm H}(8U+17\lambda)+(2U+3\lambda)(U+3\lambda)} 
\end{align}
The terms in the upper left corner of $G_j^0$ describe kinetic exchange processes where an additional hole is added to the $j_\text{eff}=1/2$ state. As usual, such processes are limited by Pauli exclusion, which generates an effective exchange interaction. The terms in the bottom right corner arise from effective Hund's coupling between a hole added to the $j_\text{eff}=3/2$ state, and the existing hole in the $j_\text{eff}=1/2$ state. 

We now discuss the derivation of the interactions for the Z$_1$ bond with $t_1=t_3=t_4=0$ as an exercise. It is convenient to rewrite the hopping matrices in the $j_\text{eff}$ basis:
\begin{align}
\mathbf{T}_{ij} = \left(\begin{array}{cc}\Theta_{ij}^{\frac{1}{2}\frac{1}{2}} & \Theta_{ij}^{\frac{1}{2}\frac{3}{2}} \\\Theta_{ij}^{\frac{3}{2}\frac{1}{2}} & \Theta_{ij}^{\frac{3}{2}\frac{3}{2}} \end{array} \right)
\end{align}
So that the resulting spin Hamiltonian is:
\begin{align}
\mathcal{H}_{\rm spin} = \sum_{ij} & \  \mathbb{A} \  \mathbf{S}_j \cdot \left( \mathbf{c}_{i,\frac{1}{2}}^\dagger \Theta_{ij}^{\frac{1}{2}\frac{1}{2}}\vec{\sigma}\Theta_{ji}^{\frac{1}{2}\frac{1}{2}}\mathbf{c}_{i,\frac{1}{2}} \right) \\ \nonumber
& \ + \mathbb{B}\ \mathbf{S}_j \cdot \left( \mathbf{c}_{i,\frac{1}{2}}^\dagger \Theta_{ij}^{\frac{1}{2}\frac{3}{2}}\vec{\tau}\Theta_{ji}^{\frac{3}{2}\frac{1}{2}}\mathbf{c}_{i,\frac{1}{2}} \right) + (i\leftrightarrow j)
\end{align}
For the pure $t_2$ limit, all hopping between $j_\text{eff}=1/2$ states vanishes, i.e. $\Theta_{ij}^{\frac{1}{2}\frac{1}{2}} = 0$. The only hopping relevant at second order is:
\begin{align}
\Theta_{ji}^{\frac{3}{2}\frac{1}{2}} = -i\sqrt{\frac{2}{3}}t_2\left(\begin{array}{cc} 0 1 \\ 0 0 \\ 0 0 \\ 1 0 \end{array} \right)
\end{align}
It is easy to show that:
\begin{align}
\Theta_{ij}^{\frac{1}{2}\frac{3}{2}} \ \tau_x \ \Theta_{ji}^{\frac{3}{2}\frac{1}{2}} = & \  0 \\ 
\Theta_{ij}^{\frac{1}{2}\frac{3}{2}} \ \tau_y \ \Theta_{ji}^{\frac{3}{2}\frac{1}{2}} = & \  0  \\
\Theta_{ij}^{\frac{1}{2}\frac{3}{2}} \ \tau_z \ \Theta_{ji}^{\frac{3}{2}\frac{1}{2}} = & \  -2 \ t_2^2  \ \sigma_z
\end{align}
Given Eq. (\ref{eqn-spin}), and summing over $(i\leftrightarrow j)$, we have:
\begin{align}
\mathcal{H}_{\rm spin} = \sum_{ij} -8 \ \mathbb{B} \ t_2^2 \  S_i^z S_j^z
\end{align}
The Ising form of the interaction arises because hopping can only occur to the $m_j = \pm 3/2$ states. The spin-flip components of the effective Hund's coupling $\vec{\tau}$ are therefore irrelevant, as only the $\tau_z$ component operates in this subspace.

Finally, we compute interactions for general hopping. Hopping integrals between sites $i,j$ are written in the $d$ basis in terms of labels ($x=d_{yz}$, $y=d_{xz}$, $z=d_{xy}$), so that:
\begin{align}
\mathbf{T}_{ij} = \left(\begin{array}{ccc}t_{xx} & t_{xy} & t_{xz} \\ t_{yx} & t_{yy} & t_{yz} \\ t_{zx} & t_{zy} & t_{zz} \end{array} \right)\otimes \mathbb{I}_{2\times 2}
\end{align}
In terms of such hopping integrals and the constants $\mathbb{A},\mathbb{B}$, the isotropic exchange constant is:
\begin{align}
J = & \ \frac{4\mathbb{A}}{27}\left\{\begin{array}{c}3(t_{xx}+t_{yy}+t_{zz})^2 -(t_{yz}-t_{zy})^2\\  -(t_{yx}-t_{xy})^2-(t_{zx}-t_{xz})^2\end{array}\right\} \\
\nonumber
& \ - \frac{4\mathbb{B}}{27}\left\{\begin{array}{c}3(t_{xx}-t_{yy})^2+3(t_{xx}-t_{zz})^2\\ +3(t_{yy}-t_{zz})^2+2(t_{xy}+t_{yx})^2\\ +2(t_{xz}+t_{zx})^2 +2(t_{yz}+t_{zy})^2 \\ +10t_{xz}t_{zx}+10t_{yz}t_{zy}+10t_{xy}t_{yx}\end{array}\right\}
\end{align}
The components of the Dzyaloshinskii-Moriya vector are $\mathbf{D}_{ij} = (D_a,D_b,D_c)$, where:
\begin{align}
D_a = & \  \frac{8\mathbb{A}}{9} \left\{\begin{array}{c}( t_{xx}+t_{yy}+t_{zz})(t_{yz}-t_{zy}) \end{array}\right\} \\ \nonumber & \ +\frac{8\mathbb{B}}{9} \left\{\begin{array}{c}(2t_{xx}-t_{yy}-t_{zz})(t_{yz}-t_{zy})\\ +3t_{xy}t_{zx} -3t_{xz}t_{yx}\end{array}\right\}
 \end{align}
\begin{align}
D_b = & \  \frac{8\mathbb{A}}{9} \left\{\begin{array}{c}( t_{xx}+t_{yy}+t_{zz})(t_{zx}-t_{xz}) \end{array}\right\} \\ \nonumber & \ +\frac{8\mathbb{B}}{9} \left\{\begin{array}{c}(2t_{yy}-t_{xx}-t_{zz})(t_{zx}-t_{xz}) \\ +3t_{xy}t_{yz} -3t_{zy}t_{yx}\end{array}\right\} 
 \end{align}
\begin{align}
D_c = & \  \frac{8\mathbb{A}}{9} \left\{\begin{array}{c}( t_{xx}+t_{yy}+t_{zz})(t_{xy}-t_{yx}) \end{array}\right\} \\ \nonumber & \ +\frac{8\mathbb{B}}{9} \left\{\begin{array}{c}(2t_{zz}-t_{xx}-t_{yy})(t_{xy}-t_{yx}) \\ +3t_{yz}t_{zx} -3t_{xz}t_{zy}\end{array}\right\} 
\end{align}
The pseudo-dipolar tensor is written:
\begin{align}
\mathbf{\Gamma}_{ij} = \left(\begin{array}{ccc}\Gamma_{aa} & \Gamma_{ab} & \Gamma_{ac} \\ \Gamma_{ab} & \Gamma_{bb} & \Gamma_{bc} \\ \Gamma_{ac} & \Gamma_{bc} & \Gamma_{cc} \end{array} \right)
\end{align}
where $\Gamma_{aa}+\Gamma_{bb}=-\Gamma_{cc}$. The diagonal terms are:
\begin{align}
\Gamma_{aa} = & \ \frac{8\mathbb{A}}{27}\left\{\begin{array}{c}2(t_{yz}-t_{zy})^2\\ -(t_{xz}-t_{zx})^2-(t_{xy}-t_{yx})^2\end{array}\right\} \\
\nonumber
& \ + \frac{4\mathbb{B}}{27}\left\{\begin{array}{c}6(t_{yy}-t_{xx})^2+6(t_{xx}-t_{zz})^2 \\ -12(t_{yy}-t_{zz})^2 \\ +5(t_{xz}-t_{zx})^2 +5(t_{xy}-t_{yx})^2 \\ -10 (t_{yz}-t_{zy})^2  \\ +4t_{yz}t_{zy}-2t_{xy}t_{yx}-2t_{xz}t_{zx}\end{array}\right\}
\end{align}
\begin{align}
\Gamma_{bb} = & \ \frac{8\mathbb{A}}{27}\left\{\begin{array}{c}2(t_{zx}-t_{xz})^2\\ -(t_{yx}-t_{xy})^2-(t_{yz}-t_{zy})^2\end{array}\right\} \\
\nonumber
& \ + \frac{4\mathbb{B}}{27}\left\{\begin{array}{c}6(t_{zz}-t_{yy})^2+6(t_{yy}-t_{xx})^2 \\ -12(t_{xx}-t_{yy})^2 \\ +5(t_{yx}-t_{xy})^2 +5(t_{yz}-t_{zy})^2 \\ -10 (t_{zx}-t_{xz})^2  \\ +4t_{zx}t_{xz}-2t_{yz}t_{zy}-2t_{yx}t_{xy}\end{array}\right\}\end{align}
The off-diagonal terms of the pseudo-dipolar tensor are:
\begin{align}
\Gamma_{ab} = & \ \frac{8\mathbb{A}}{9}\left\{\begin{array}{c}(t_{xz}-t_{zx})(t_{zy}-t_{yz})\end{array}\right\}\\
\nonumber & \ +\frac{4\mathbb{B}}{9}\left\{ \begin{array}{c}3(t_{xx}+t_{yy}-2t_{zz})(t_{yx}+t_{xy})\\ +5(t_{xz}+t_{zx})(t_{yz}+t_{zy})\\ -t_{xz}t_{zy}-t_{yz}t_{zx}\end{array}\right\}\end{align}

\begin{align}
\Gamma_{bc} = & \ \frac{8\mathbb{A}}{9}\left\{\begin{array}{c}(t_{yx}-t_{xy})(t_{xz}-t_{zx})\end{array}\right\}\\
\nonumber & \ +\frac{4\mathbb{B}}{9}\left\{ \begin{array}{c}3(t_{yy}+t_{zz}-2t_{xx})(t_{zy}+t_{yz})\\ +5(t_{yx}+t_{xy})(t_{zx}+t_{xz})\\ -t_{yx}t_{xy}-t_{zx}t_{xy}\end{array}\right\}\end{align}

\begin{align}
\Gamma_{ac} = & \ \frac{8\mathbb{A}}{9}\left\{\begin{array}{c}(t_{zy}-t_{yz})(t_{yx}-t_{xy})\end{array}\right\}\\
\nonumber & \ +\frac{4\mathbb{B}}{9}\left\{ \begin{array}{c}3(t_{zz}+t_{xx}-2t_{yy})(t_{xz}+t_{zx})\\ +5(t_{zy}+t_{yz})(t_{xy}+t_{yx})\\ -t_{zy}t_{yz}-t_{xy}t_{yz}\end{array}\right\}\end{align}

\bibliography{entire}
\end{document}